\documentclass[a4paper,11pt]{article}
\usepackage{jinstpub} 
\usepackage{lmodern}
\usepackage{lineno}
\usepackage{float} 
\usepackage{booktabs}
\usepackage{multirow}
\usepackage{mathtools} 
\usepackage{wasysym} 
\graphicspath{{figures/}{}} 

\title{\boldmath Design, fabrication and testing of Al/\textit{p}-Si Schottky and \textit{pn} junctions for radiation studies}

\author[a,*]{E. Giulio Villani\note{Corresponding author.}}
\author[b,*]{Dengfeng Zhang\note{Corresponding author.}}
\author[c]{Adnan Malik} 
\author[b]{Trevor Vickey}
\author[d,e]{Yebo Chen}
\author[d]{Matthew G. Kurth}
\author[d]{Peilian Liu}
\author[h]{Hongbo Zhu}
\author[f]{Thomas Koffas}
\author[f]{Christoph Thomas Klein}
\author[g]{Robert Vandusen}
\author[g]{Rodney Aiton}
\author[g]{Angela Mccormick}
\author[g]{Garry Tarr}
\affiliation[a]{STFC Rutherford Appleton Laboratory Particle Physics Department, \\Harwell, Didcot, OX110QX, UK}
\affiliation[b]{Department of Physics and Astronomy, The University of Sheffield, \\Western Bank, Sheffield, S10 2TN, UK}
\affiliation[c]{STFC Rutherford Appleton Laboratory, Innovations Technology Access Centre, \\Harwell, Didcot, OX110QX, UK}
\affiliation[d]{Institute of High Energy Physics, Chinese Academy of Sciences, \\Beijing 100049, China}
\affiliation[e]{University of Chinese Academy of Sciences, \\Beijing, 100049, China}
\affiliation[f]{Department of Physics, Carleton University, \\Ottawa, Ontario, K1S 5B6, Canada}
\affiliation[g]{Department of Electronics, Carleton University, \\Ottawa, Ontario, K1S 5B6, Canada}
\affiliation[h]{School of Physics, Zhejiang University, \\Hangzhou, 310058, China}

\emailAdd{giulio.villani@stfc.ac.uk}
\emailAdd{dengfeng.zhang@cern.ch}

\abstract{Strip and pixels sensors, fabricated on high resistivity silicon substrate, normally of \textit{p}-type, are used in detectors for High Energy Physics (HEP) typically in a hybrid detector assembly. 
Furthermore, and owing to their inherent advantages over hybrid sensors, Monolithic Active Pixel Sensors (MAPS) fabricated in CMOS technology have been increasingly implemented in HEP experiments. In all cases, their use in higher radiation areas (HL-LHC and beyond) will require options to improve their radiation hardness and time resolution. 
These aspects demand a deep understanding of their radiation damage and reliable models to predict their behaviours at high fluences. As a first step, we fabricated several Schottky and \textit{n}-on-\textit{p} diodes, to allow a comparison of results and provide a backup solution for test devices, on 6 or 4-inch \textit{p}-type silicon wafers with 50 $\mu$m epitaxial thickness and of doping concentration as they are normally used in HEP detectors and CMOS MAPS devices. 
In this paper, details of the design and fabrication process, along with test results of the fabricated devices before irradiation, will be provided. Additional test results on irradiated devices will be provided in subsequent publications.}


\arxivnumber{} 

\begin{document}

\maketitle
\flushbottom

\section{Introduction}
\label{section:introduction}
In the High Energy Physics (HEP) community the topic of radiation hardness of detectors and related electronics has been and will continue to be, of crucial importance. The integrated luminosity of colliders has seen an almost constant increase by a factor of 10 per decade over the last 50 years~\cite{schmickler2022colliders}. As a consequence, the required radiation hardness of the sensors and electronics for the operation on the modern HEP detectors, like ATLAS and CMS on Large Hadron Collider (LHC) at CERN, has increased accordingly~\cite{Dawson:2764325}.

The technology usually implemented on HEP is hybrid detectors making use of high resistivity silicon~\cite{Lutz:1999wg}, normally \textit{p}-type.
Another technology, which has received much attention
owing to its inherent advantages over hybrid sensors, is the Monolithic Active Pixel Sensors (MAPS) fabricated in CMOS technology~\cite{TURCHETTA2001677}, which has already been applied in several HEP experiments~\cite{MAGER2016434,CONTIN20167,Vilella_2018}.
Typically, the silicon used for the CMOS MAPS fabrication is \textit{p}-type and consists of a medium-high resistivity epitaxially grown layer on top of a low doping substrate~\cite{SNOEYS2013125}.
Significant performance degradation is usually observed after fluences of $1\times 10^{13}$ n\textsubscript{eq}/cm\textsuperscript{2}~\cite{Dawson:2764325}, as a result of bulk damage to the crystal structure. 
Despite various modifications having been proposed and implemented to improve radiation hardness~\cite{1462422,SNOEYS201790}, a wider use of CMOS MAPS in higher radiation areas (e.g., the High-Luminosity LHC and beyond) will require an even deeper understanding of the mechanism of radiation bulk damage and mitigation methods, along with reliable models to predict their behaviors after high fluences.
To investigate radiation bulk damage,  we fabricated a number of Schottky and \textit{n}-on-\textit{p} junction diodes on \textit{p}-type epitaxial silicon wafers, with doping concentrations typically used in CMOS MAPS devices. The purpose is to gain a deeper understanding of radiation damage in such structures with a view to develop reliable damage models that can be implemented in TCAD device simulators, like Synopsys TCAD ~\cite{Synopsys}.

In this paper, Section \ref{section:schottky diodes and pn} will give a description of the Schottky and \textit{pn} junction diode device layouts. 
In Section \ref{section:details of device fabrication}, a description of the fabrication process of the devices will be given.
Section \ref{section:test results} will present test results, including Current-Voltage (IV), Capacitance-Voltage (CV), Charge Collection Efficiency (CCE) and defects obtained using the Deep-Level Transient Spectroscopy (DLTS) technique. Finally, in Section \ref{sectioin:conclusion and plan}, preliminary conclusions and a description of the next steps for this project will be given.

\section{Schottky diode and \textit{pn} junction design}
\label{section:schottky diodes and pn}
In this research project several layouts of different device sizes were used. 
Bigger devices were designed to facilitate the measurements of capacitance and leakage current (CV and IV),
whilst smaller devices, of enhanced signal-to-noise ratio (S/N) due to their smaller capacitance, were designed to investigate the charge collection efficiency (CCE) when injecting charge via minimum ionizing particle (MIP) or laser pulse.
The Schottky diodes were fabricated on 6$''$ silicon wafers at Innovations Technology Access Centre (ITAC) at Rutherford Appleton Laboratory (RAL). The \textit{pn} junctions were fabricated at the Carleton University MicroFabrication Facility (CUMFF) on 4$''$ silicon wafers sized down using laser dicing.
And these silicon wafers were purchased from PlutoSemi~\cite{PlutoSemi}.

\subsection{Schottky diode device layout}
\label{subsection:schottky diode device layout}
\begin{figure}[htbp]
\centering
\includegraphics[width=0.4\textwidth]{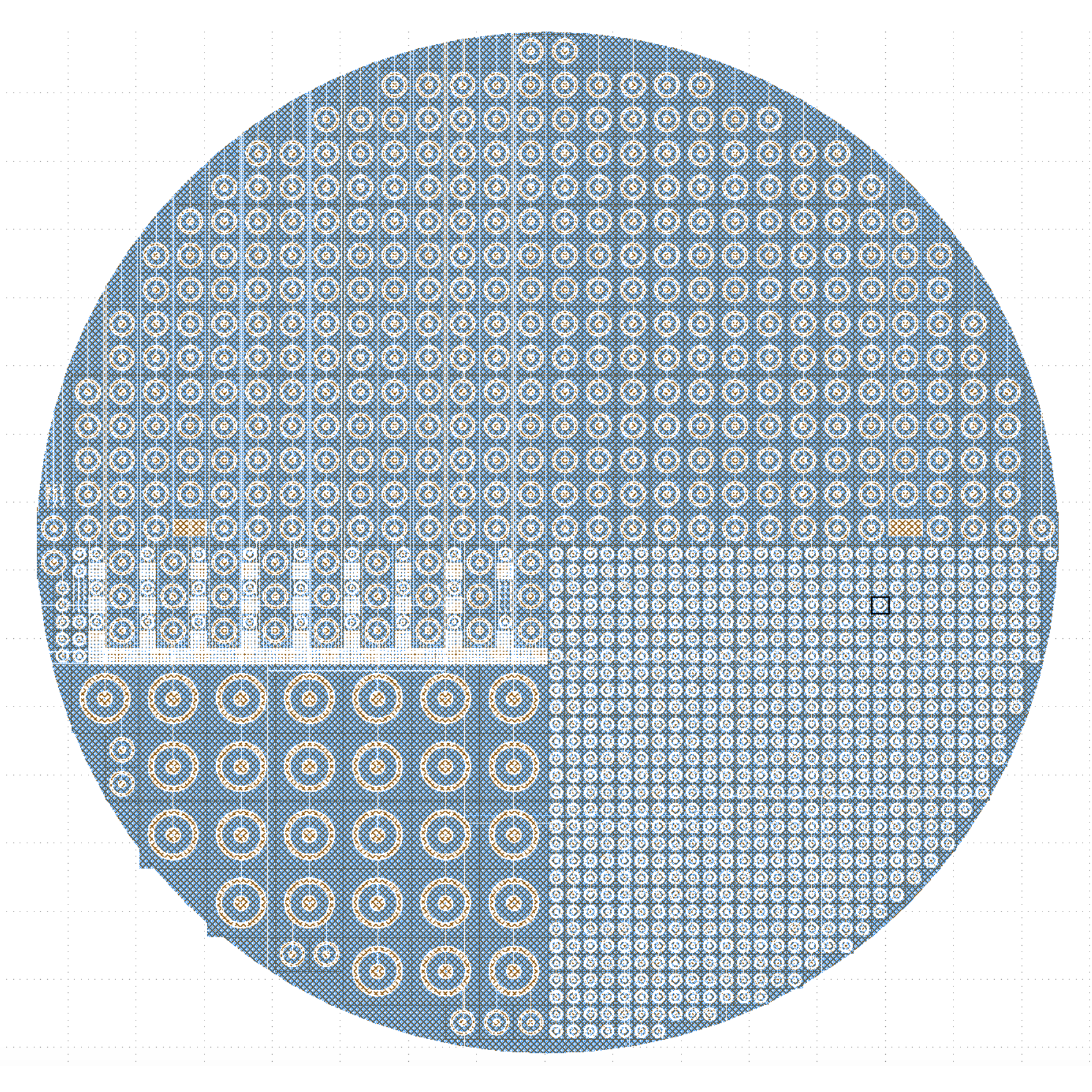}
\includegraphics[width=0.4\textwidth]{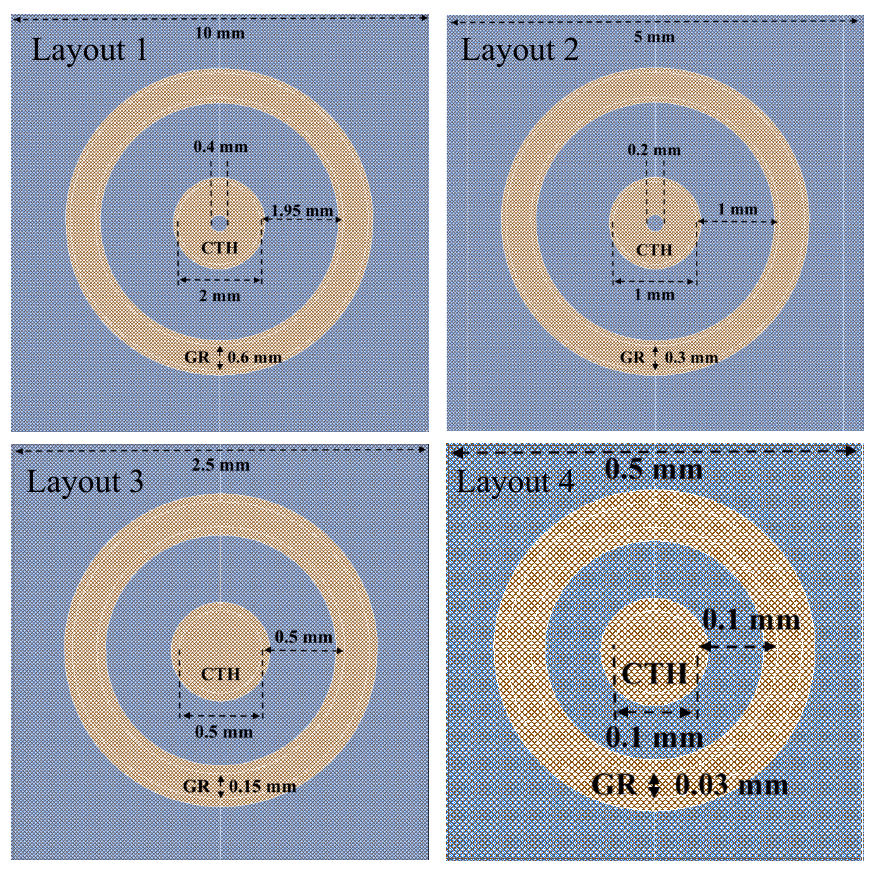}
\caption{Left: The Schottky device map on the 6$''$ wafer. Right: Four layouts of the Schottky diode devices.}
\label{fig:schottkylayout}
\end{figure}

The full wafer map along with the four device layouts are shown in Figure \ref{fig:schottkylayout}. As the fabrication process of the Schottky diode is relatively simple, only two sets of 7$''$ masks were used, one for the metal and the other for the oxide. 
The different device layouts implemented differ in the cathode diameter and guard ring size as reported in Table \ref{table:schottkydiode_layout}, there are four layouts in total (Layout 1-T1, Layout 2-T2, Layout 3-T3, Layout 4-T4). 
The layout 1 and layout 2 devices have a cathode of 2 mm and 1 mm diameter respectively, with a central hole in the metal to allow charge injection via a laser pulse for the charge colleciton investigation. 
The other two smaller devices have cathodes of 0.5 mm and 0.1 mm diameter respectively, and do not have a central hole. In all devices, the metal cathode slightly overlaps with the dielectric to reduce the effect of early breakdown~\cite{BECK2001183}.

\begin{table}[htbp]
\centering
\caption{Layout details of Schottky devices.
\label{table:schottkydiode_layout}}
\begin{tabular}{lcccc}
\toprule
Size [mm] &Layout 1 (T1) & Layout 2 (T2) & Layout 3 (T3) & Layout 4 (T4)\\
\midrule
Cathode $\diameter$\protect\footnotemark  & 2 &1 &0.5 &0.1\\
Guard ring width   & 0.6 &0.3 &0.15 &0.03\\
Central hole $\diameter$  & 0.4 &0.2 &N/A &N/A\\
\bottomrule
\end{tabular}
\end{table}

\subsection{\textit{pn} junction device layout}
\label{subsection:pN junction device layout}
\begin{figure}[htbp]
\centering
\includegraphics[width=.6\textwidth]{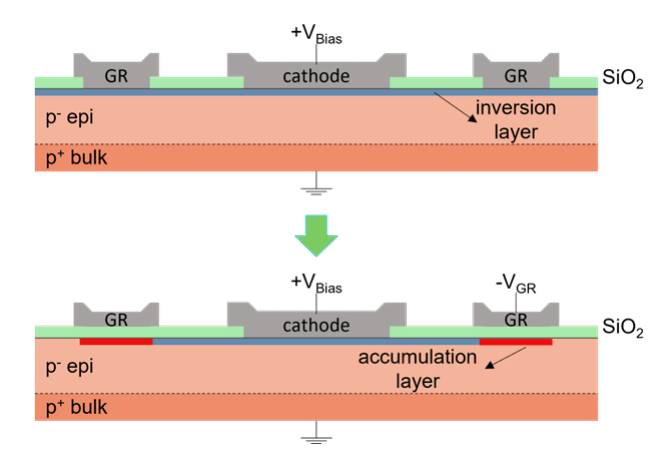}
\caption{Schematic of the \textit{pn} junction devices with REG-GR (top) and ISO-GR (bottom) guard ring.}
\label{fig:pnlayout}
\end{figure}

The \textit{pn} junction devices share the same geometry and layout as the Schottky diodes of Figure \ref{fig:schottkylayout}, the only difference being in the guard ring implementation. The metal layer of the guard ring of the \textit{pn} junction device either contacts the epitaxial layer directly (non-isolated guard ring/REG-GR) or is isolated from it by a layer of oxide (isolated guard ring/ISO-GR) to mitigate the formation of an inversion layer at the Si/SiO\textsubscript{2} interface. Cross-sections of the two types of the \textit{pn} junction devices are shown in  Figure \ref{fig:pnlayout}. An additional device flavour includes a \textit{p}-stop implemented in the epitaxial layer under the guard ring.

\footnotetext{$\diameter$: the diameter of the cathode or the central hole.}

\section{Details of device fabrication}
\label{section:details of device fabrication}
For this project, a total of 125 silicon wafers were purchased from PlutoSemi~\cite{PlutoSemi}, consisting of 5 groups of 25 wafers each. All wafers have a 50 $\mu$m thick epitaxial \textit{p}-type layer.
The 5 groups of wafers have epitaxial resistivity of 1 k$\Omega$/cm, 100 $\Omega$/cm, 10 $\Omega$/cm, 1 $\Omega$/cm and 0.1 $\Omega$/cm respectively. 
The epitaxial layer is grown on a \textit{p}-type substrate with a thickness of 575 $\mu$m and resistivity 0.005 $\Omega$/cm. Subsets of the wafers were used for the fabrication of the Schottky diodes and \textit{pn} junctions. The fabrication of the Schottky diode devices was performed by ITAC at Rutherford Appleton Laboratory, the \textit{pn} junction devices were fabricated at CUMFF at Carleton University.

\subsection{Fabrication of the Schottky diode devices}
\label{subsection:fabrication of the schottky diode devices}
\begin{figure}[htbp]
\centering
\includegraphics[width=0.4\textwidth]{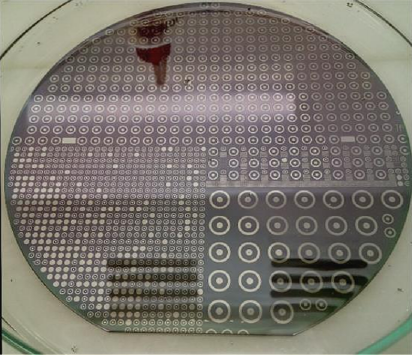}
\includegraphics[width=0.4\textwidth]{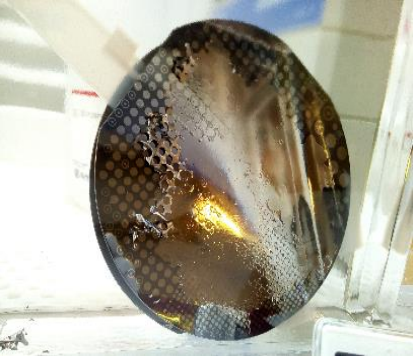}
\includegraphics[width=0.4\textwidth]{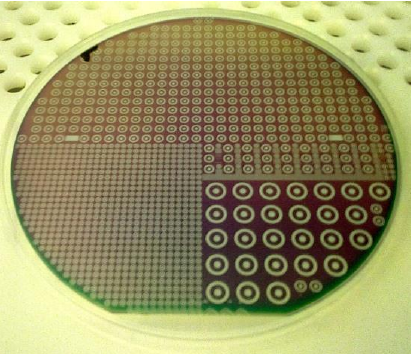}
\includegraphics[width=0.4\textwidth]{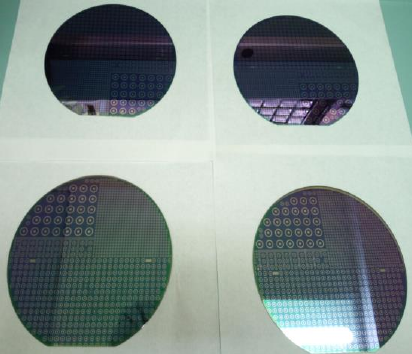}
\caption{Top Left: Wafer during resist ashing. Top Right: Example of the lift-off process, showing the top metal peeling off the surface of the wafer. Bottom Left: The first completed wafer with Schottky diode devices on a high resistivity epi layer(doping concentration: $10^{13}$ cm\textsuperscript{-3}). Bottom Right: The first four completed wafers, two on high resistivity epi layer (doping concentration: $10^{13}$ cm\textsuperscript{-3}) and the other two on medium resistivity epi layer (doping concentration: $10^{14}-10^{15}$ cm\textsuperscript{-3}).}
\label{fig:schottkywaferfabexamples}
\end{figure}

\begin{figure}[htbp]
\centering
\includegraphics[width=0.4\textwidth]{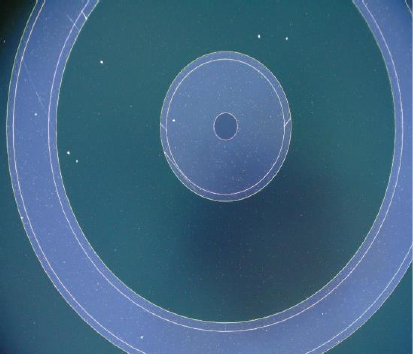}
\includegraphics[width=0.4\textwidth]{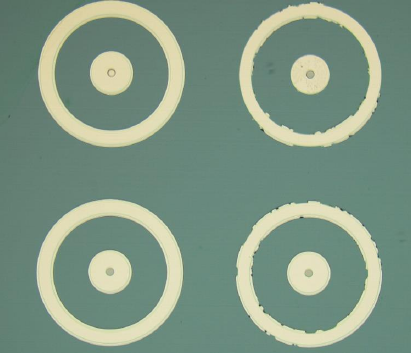}
\caption{Left: Microphotographs shows the fabricated Schottky diodes after the correct lift-off process. Right: Results of non-uniform  metal erosion following attempts to increase the process speed by increasing the bath temperature. The optimal temperature, which avoided the non-uniform rate of lift-off, was found to be 40 $^{\circ}$C, as reported in Appendix \ref{appendix:device fabrication process}.}
\label{fig:schottkydiodeexamples}
\end{figure}

For the fabrication of Schottky diodes, only two masks were required.
Full technical details of the fabrication, along with synopsys of the manufacturing process flow, are reported in Appendix \ref{appendix:device fabrication process}.
Examples of final fabricated devices and during the various stages of fabrication are shown in Figures \ref{fig:schottkywaferfabexamples} and \ref{fig:schottkydiodeexamples}.

\subsection{Fabrication of the \textit{pn} junction diode devices}
\label{subsection:fabrication of the pn junction diode devices}

\begin{figure}[htbp]
\centering
\includegraphics[width=0.4\textwidth]{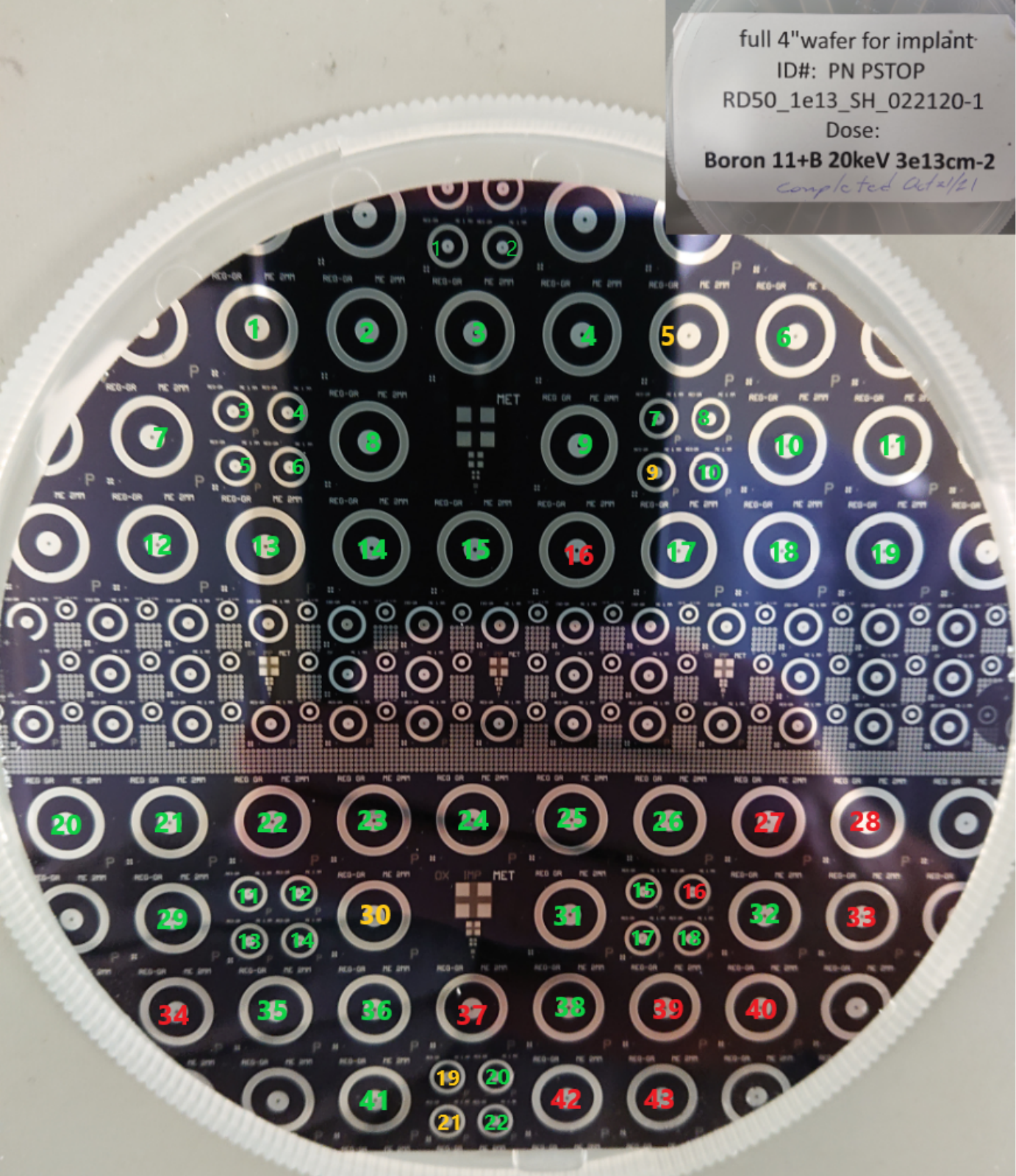}
\includegraphics[width=0.4\textwidth]{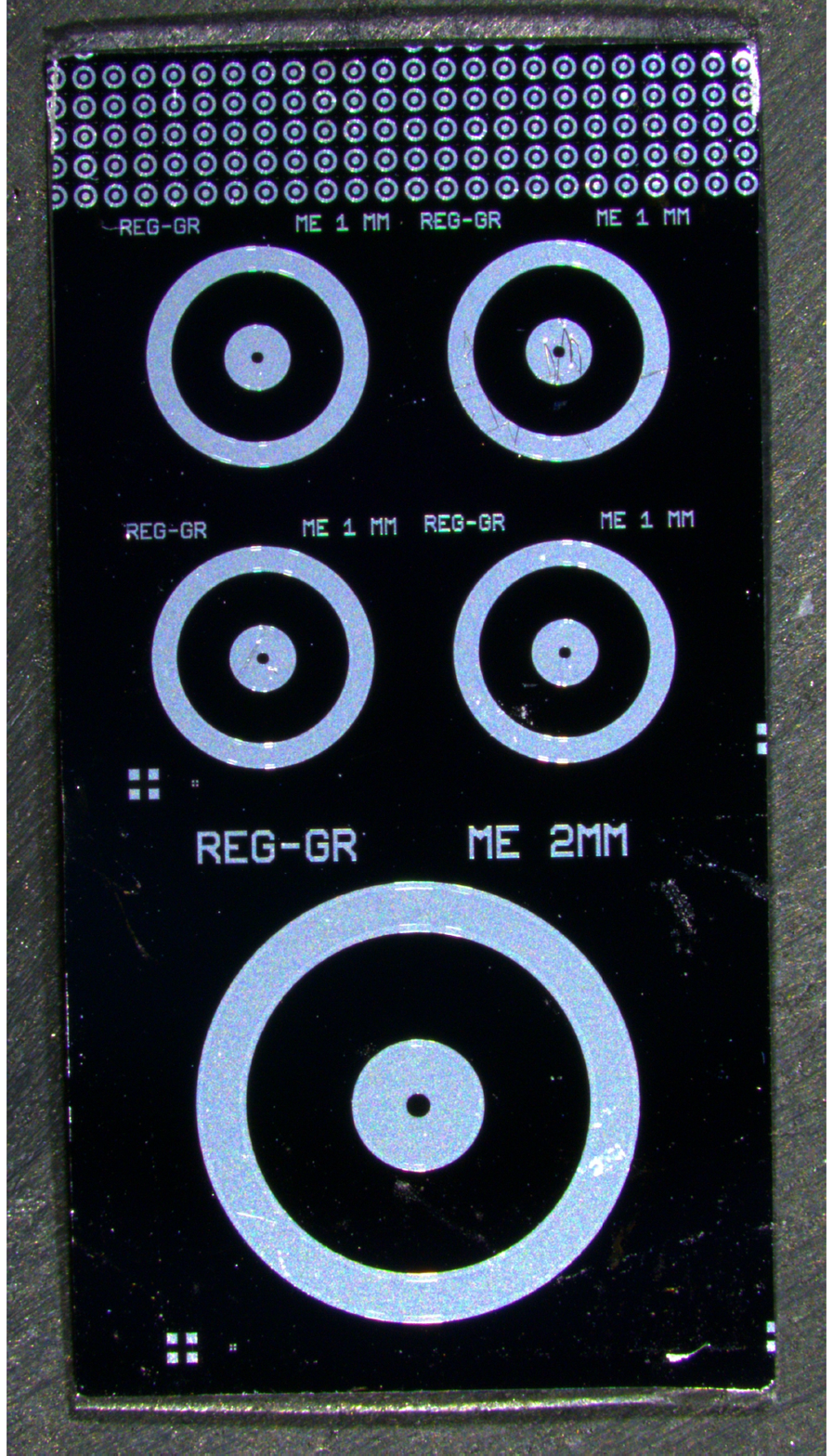}
\caption{Left: Processed \textit{pn} junction wafer. Right: Diced \textit{pn} junction die.}
\label{fig:pnjunctionwafer}
\end{figure}

The fabrication of \textit{pn} junction diode required two masks of smaller size, as the fabrication took place on 4$''$ wafers. The layout of the individual devices was the same as used for the Schottky diodes. Full technical details of the fabrication are reported in Appendix \ref{appendix:device fabrication process}.
Figure \ref{fig:pnjunctionwafer} shows an example of a processed \textit{pn} junction wafer and diced device.

\section{Test results}
\label{section:test results}

The Current-Voltage (IV) and Capacitance-Voltage (CV) characteristics of the fabricated Schottky diodes and \textit{pn} junctions were measured independently at Rutherford Appleton Laboratory (RAL), Oxford Physics Microstructure Detector (OPMD) laboratory and at Carleton University.
The charge collection efficiencies were measured at RAL. 
Defect characteristics were measured at Carleton University using a Deep-Level Transient Spectroscopy (DLTS) setup.

\subsection{Test setups}
\label{subsection:test setups}

\subsubsection*{Test setups at RAL and OPMD}
\label{subsectioin:test setups at ral and opmd}

\begin{figure}[htbp]
\centering
\includegraphics[width=0.5\textwidth]{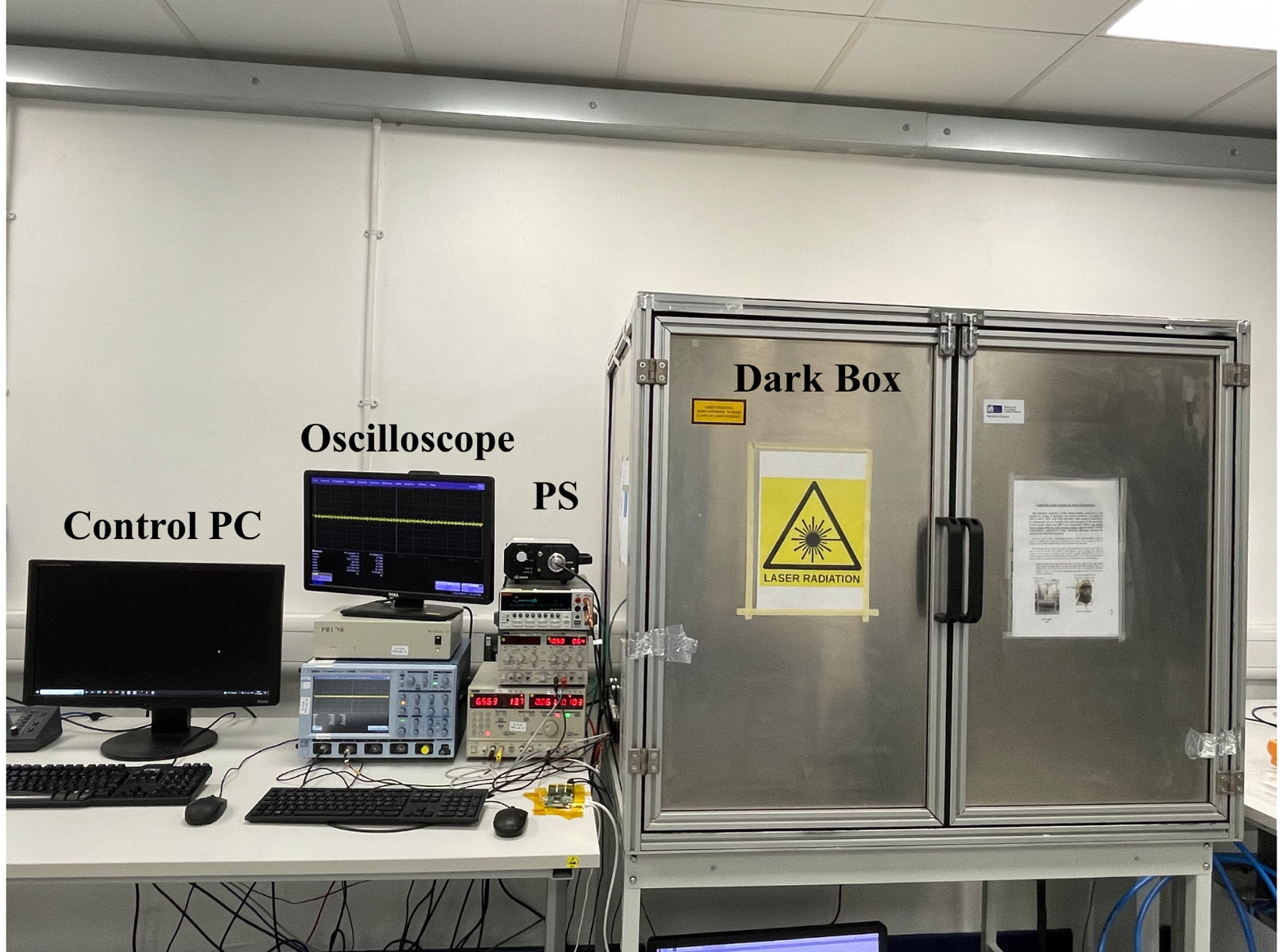}
\includegraphics[width=0.3\textwidth]{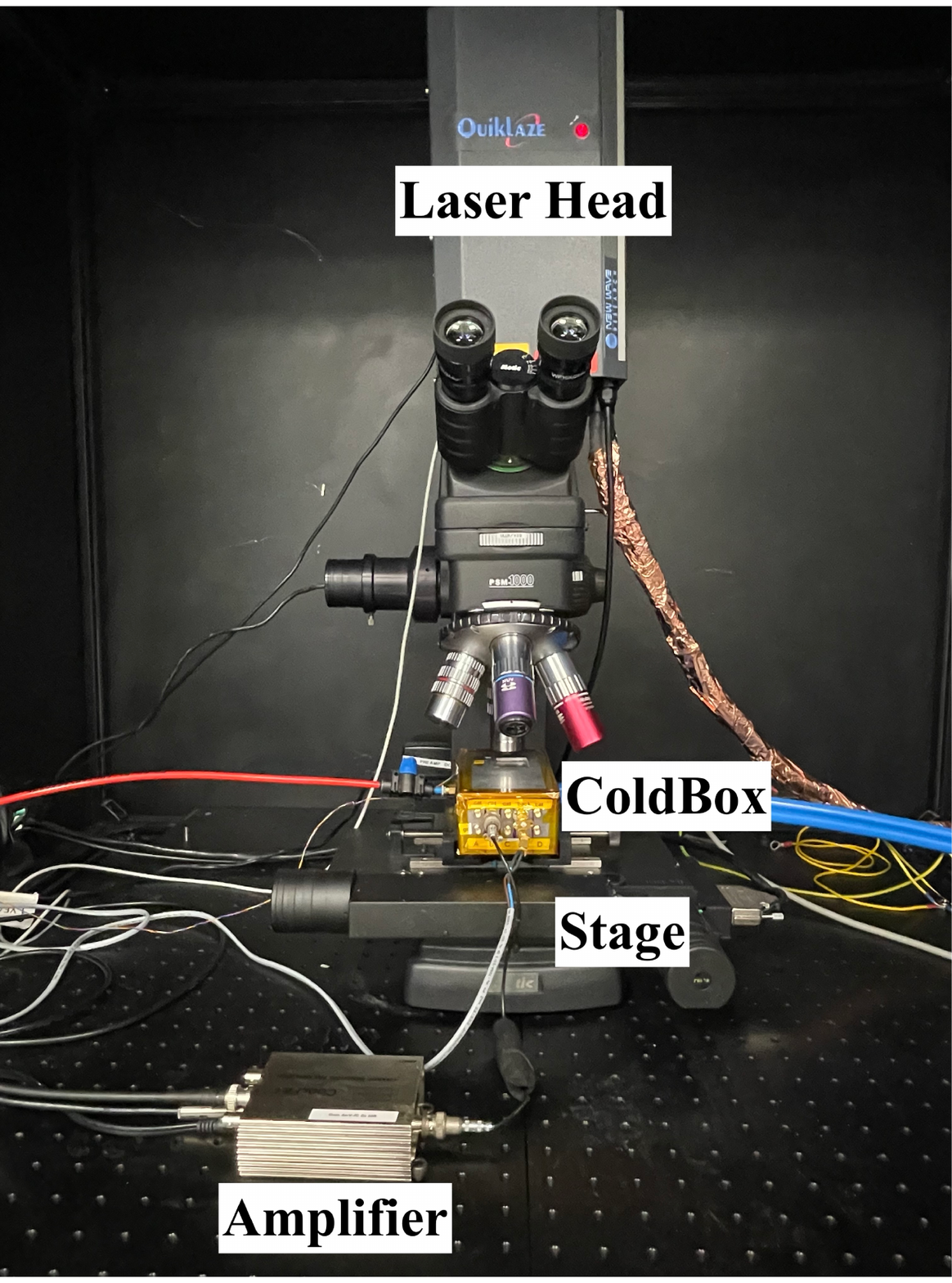}
\caption{Setup for the charge collection test at RAL. Left: Dark box housing the Trilite Laser. Right: Detail of the Trilite Laser with a cold box on the stage and the charge amplifier.}
\label{fig:RALLaserSetup}
\end{figure}

\begin{figure}[htbp]
\centering
\includegraphics[width=.6\textwidth]{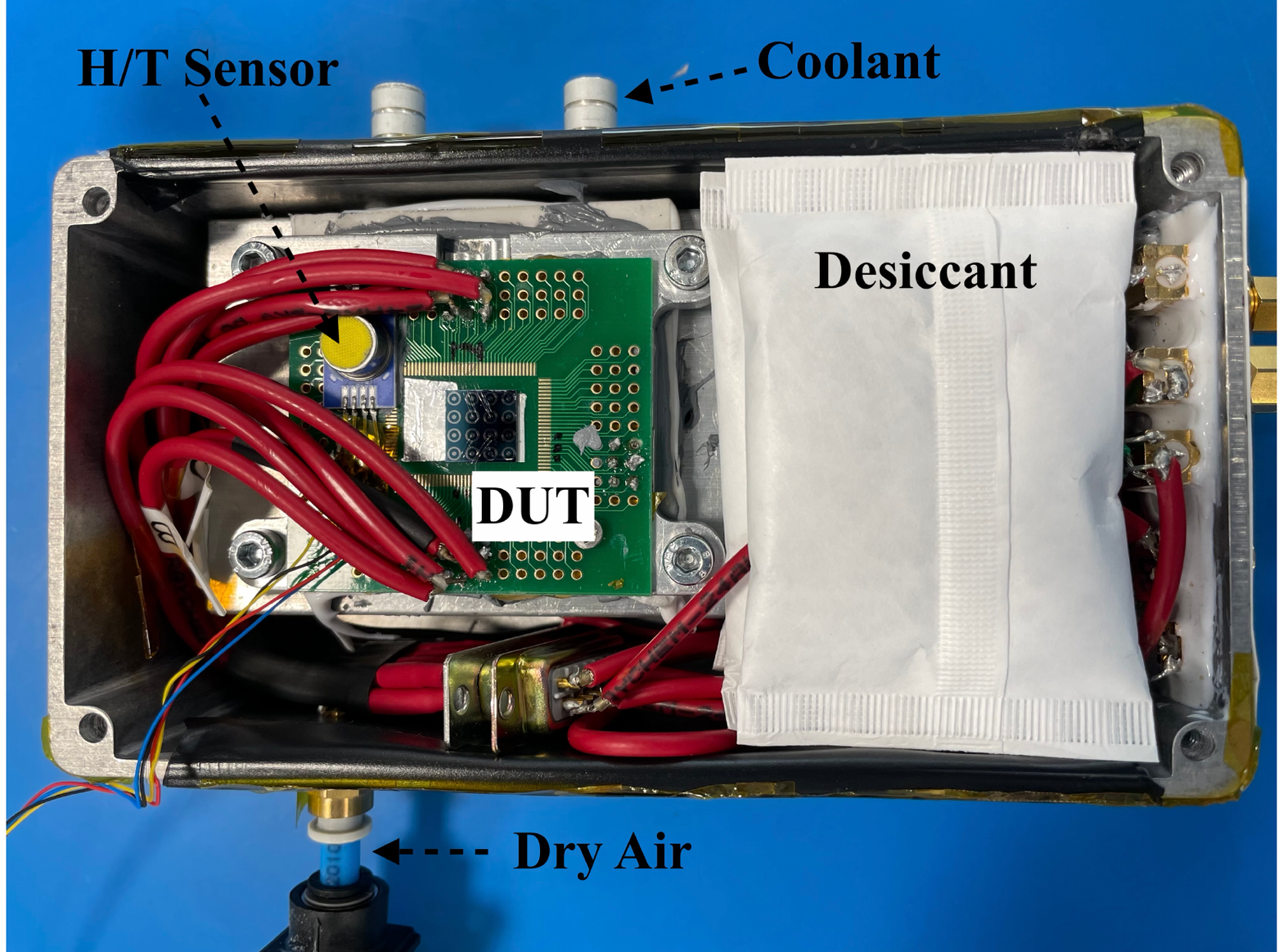}
\caption{Coldbox housing the Schottky diode and \textit{pn} junction devices for test. H/T means Humidity and Temperature.}
\label{fig:RALColdbox}
\end{figure}

The setup for the measurement of IV and CV characteristics at RAL consists of a Keithley CS4200 Semiconductor analyzer coupled to a Micromanipulator probe station and Keithley Power interface 8020. 
The setup for the IV characteristics test at OPMD consists of a Keithley 4200A-SCS Semiconductor parameter analyzer coupled to a SEMIPROBE probe station (SA-12), Keithley interface 8020 and Keithley 2657A High Power System SourceMeter. 
The devices were tested at room temperature ($\sim$20 $^{\circ}$C) in a clean room both at RAL and OPMD. 
The charge collection efficiency was measured using a custom infrared (IR) laser injection system at RAL, built around a modified QuickLaze Trilite Laser. 
Figure \ref{fig:RALLaserSetup} shows the whole setup consisting of a control PC, a LeCroy WaveRunner 6100A Oscilloscope, a Keithley 2410 HV Power Supply and a dark box where the laser header, stage, Coldbox, Amptek A250CF amplifier with a gain of 4 mV/fC are housed. 
The device under test was placed in the ColdBox, shown in Figure \ref{fig:RALColdbox}, which is supplied with dry air and uses a chiller together with a Peltier element to adjust the temperature and humidity.
The lowest temperature used for the test is $-$20 $^{\circ}$C, with relative humidity controlled down to 0.1\%. 
The IR laser beam ($\lambda$ = 1064 nm) is shaped, via electrically controlled collimators, to a rectangle of $\sim$5 $\mu$m$\times$50 $\mu$m and the beam energy was set to 23.44$\pm$2.2 pJ. A custom-written LabVIEW program was developed to perfom the full automatic tests and 1-D and 2-D scans of the DUT.

\subsubsection*{Test setups at Carleton University}
\label{subsection:test setups at carleton}
\begin{figure}[htbp]
\centering
\includegraphics[width=0.5\textwidth]{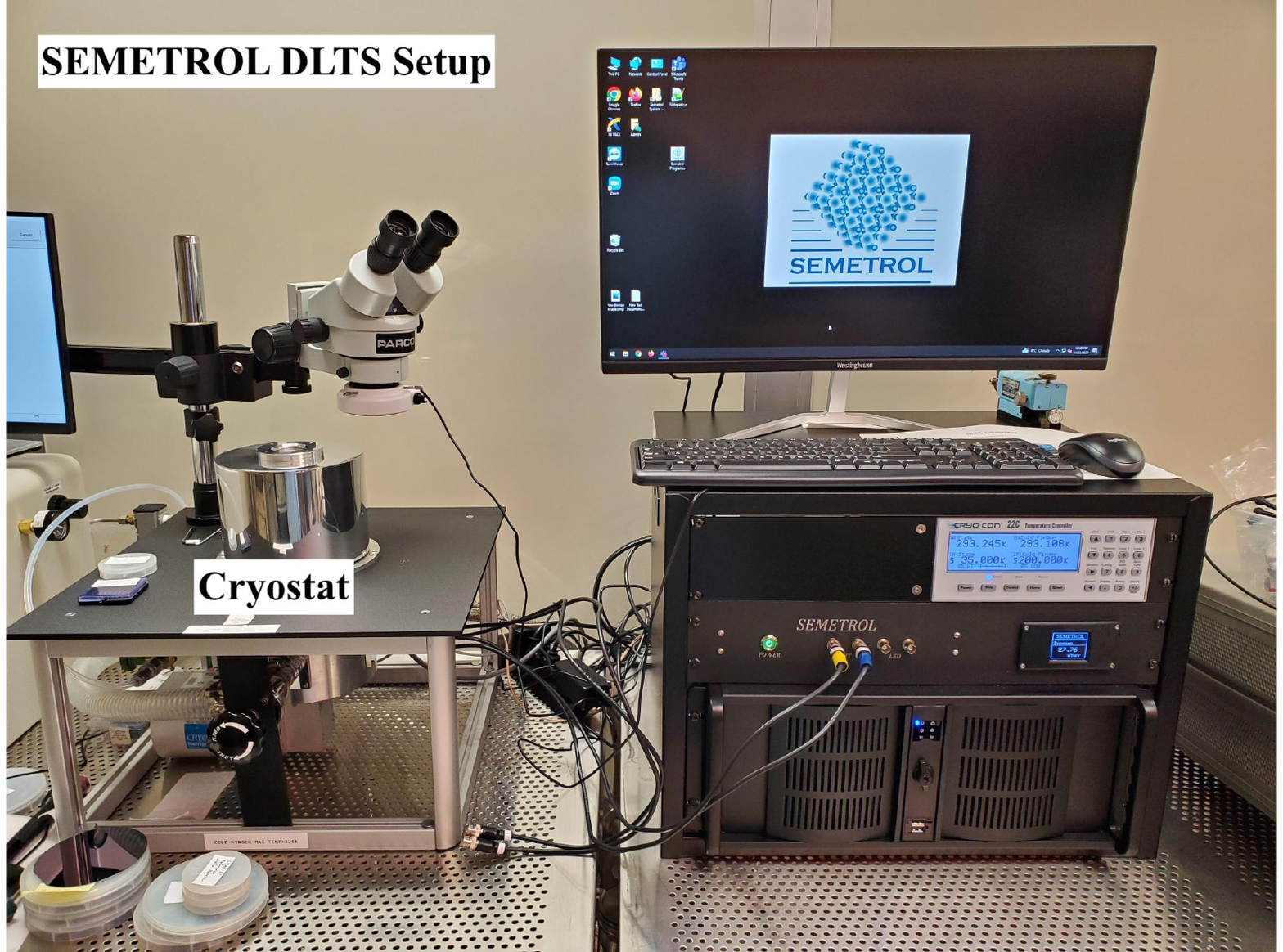}
\includegraphics[width=0.3\textwidth]{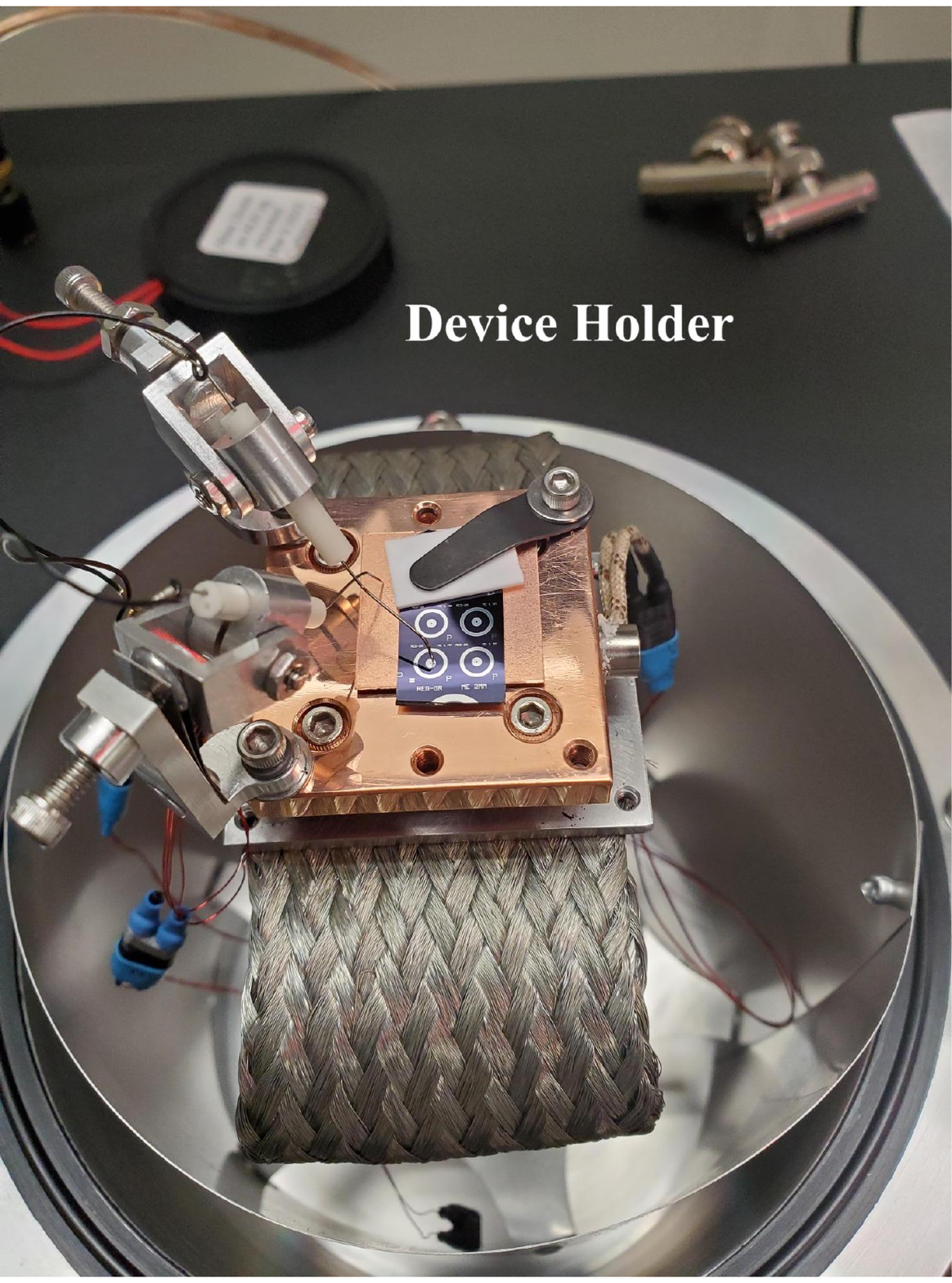}
\caption{SEMETROL DLTS system at Carleton for the measurement of the defects.}
\label{fig:CarletonDLTSSetup}
\end{figure}

For the IV characteristics measurements at Carleton, a Keithley 2410 HV PS and 6517 SMUs were used for the biasing of the device and current measurements. 
A WayneKerr 6440B LCR meter was used for the CV measurement. 
All instruments are coupled to a SemiProbe PS4L automatic probe system which allows measuring the devices on wafers after dicing.
Defect characteristics were measured using a SEMETROL DLTS system~\cite{SEMETROL}, shown in Figure \ref{fig:CarletonDLTSSetup}. 
This system uses a sample holder connected to a liquid helium cryostat, allowing it to lower the sample temperature to below 40 K or heat the sample above room temperature. 
Aside from standard DLTS measurements~\cite{10.1063/1.1663719} and temperature-dependent IV and CV scans, the SEMETROL DLTS system can also employ a wide range of spectroscopic measurements to characterise charge traps: current DLTS (I-DLTS) to analyse traps based on transients in diode current rather than capacitance; double-pulse DLTS (DDLTS) to determine the trap spatial profile and to determine the field dependence of the trap by varying the filling pulse; optical DLTS where the electrical filling pulse is replaced by an optical filling pulse from a pulsed LED; Thermal Admittance Spectroscopy (TAS) more appropriate for samples with high resistivity~\cite{BARBOLLA1992285}.

\subsection{IV test results}
\label{subsection:iv test results}

\subsubsection*{Schottky diode}
\label{subsubsection:schottky diode-iv}
\begin{figure}[htbp]
\centering
\includegraphics[width=0.4\textwidth]{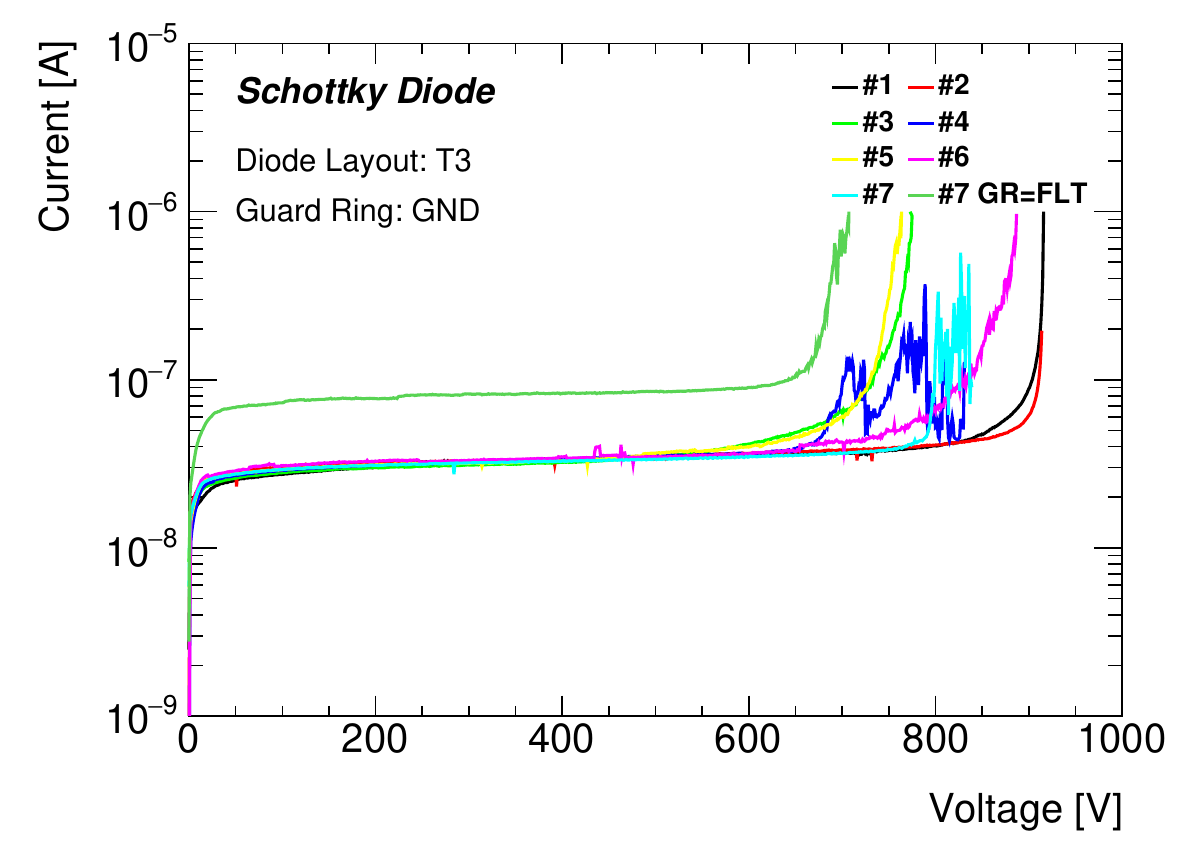}
\includegraphics[width=0.4\textwidth]{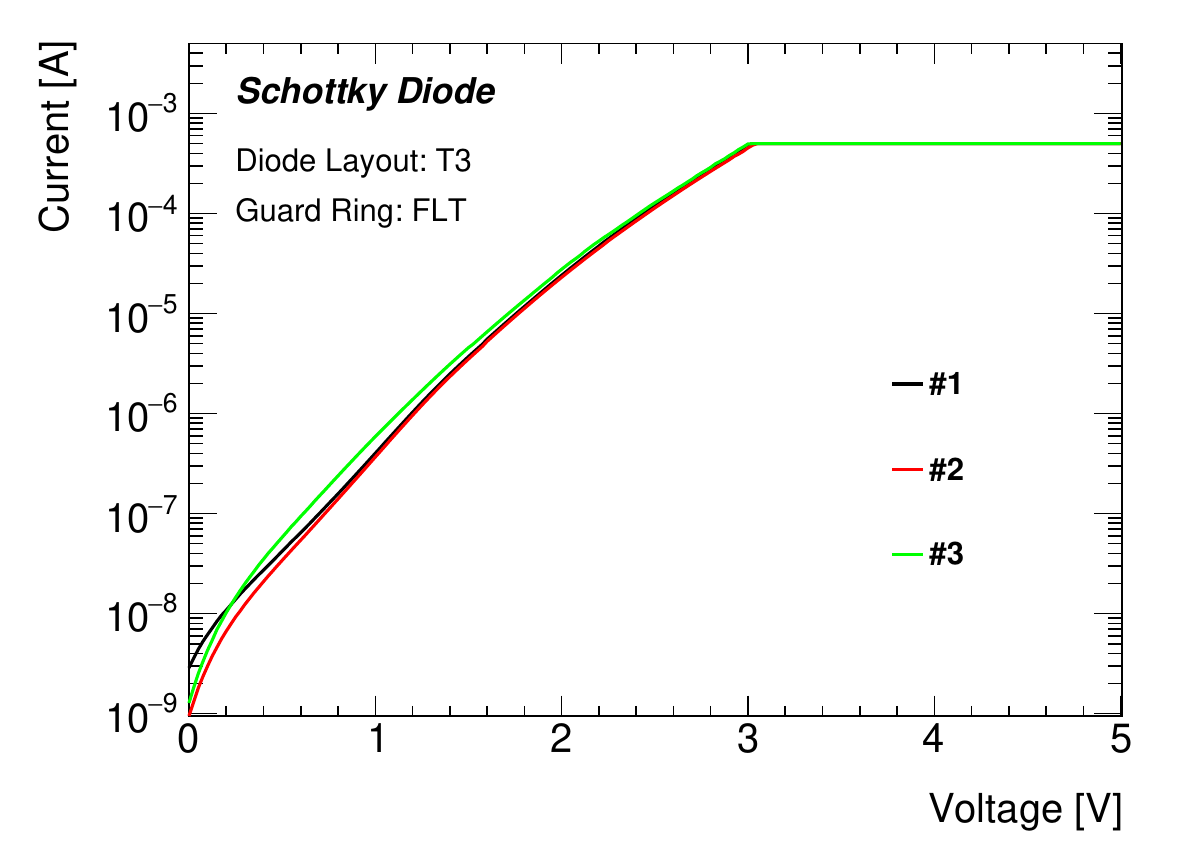}
\caption{IV characteristics of Schottky diodes of Layout 3 measured at room temperature (T = 20 $^{\circ}$C). Left: Reverse IV. Right: Forward IV.}
\label{fig:iv_schottkydiode_t3}
\end{figure}

\begin{figure}[htbp]
\centering
\includegraphics[width=0.4\textwidth]{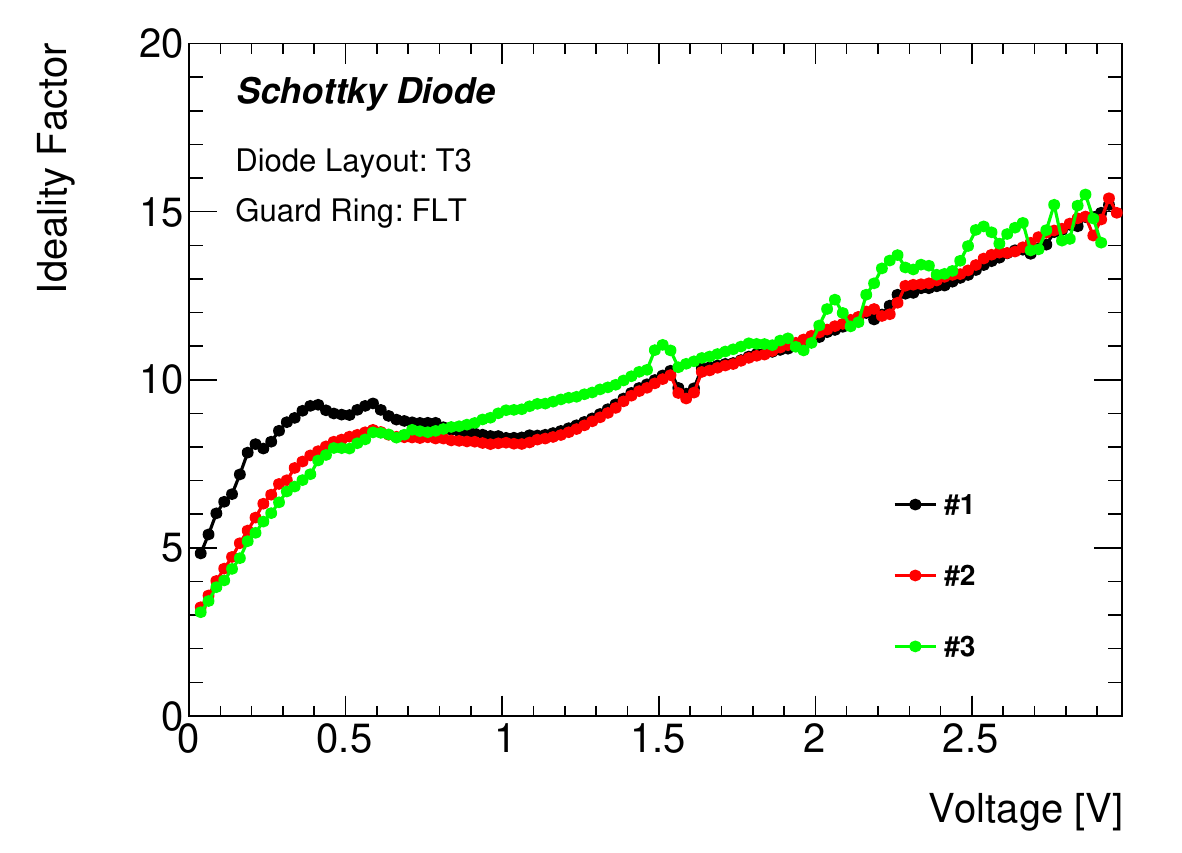}
\includegraphics[width=0.4\textwidth]{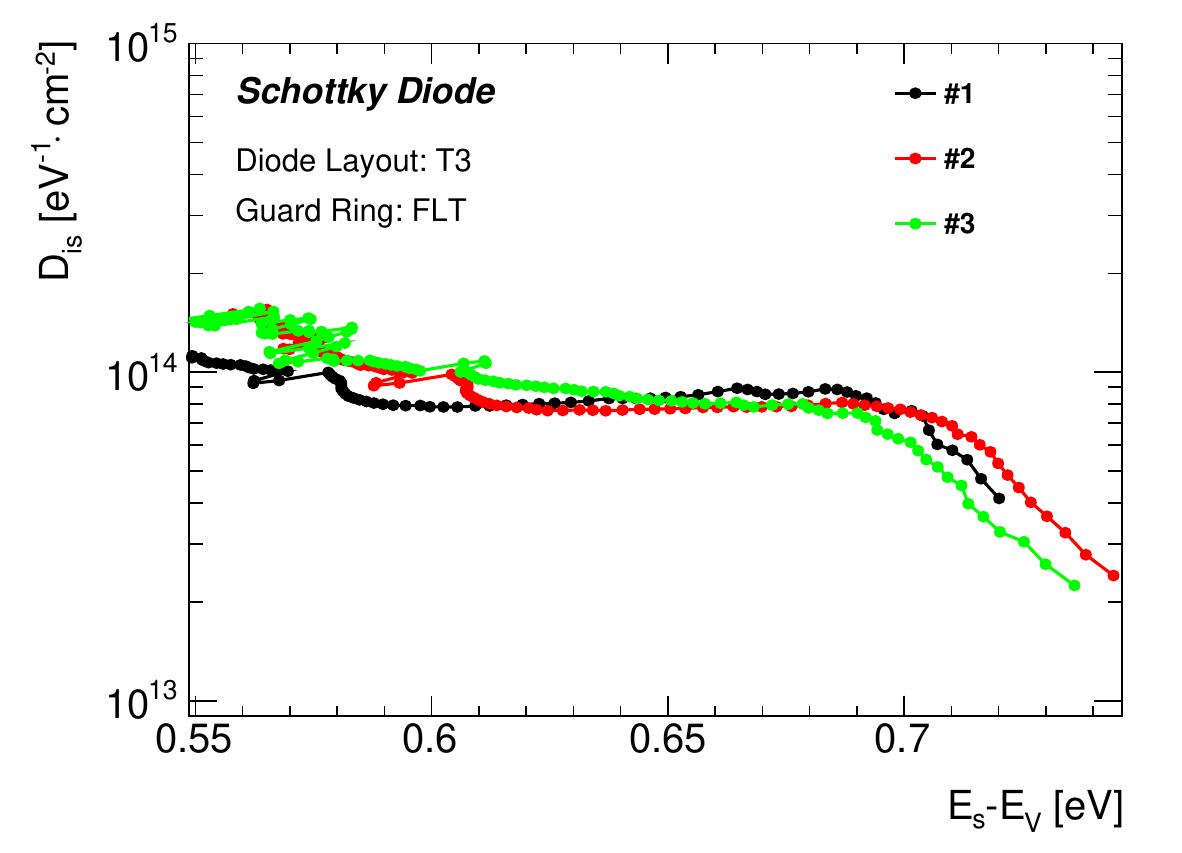}
\caption{Left: The ideality factors $n(V)$. Right: The energy distribution proﬁle of interface state densities $D_{is}(E)$.}
\label{fig:schottkydiode_idealityfactor_dis}
\end{figure}

The IV characteristics, in reverse and forward biasing mode at T = 20 $^{\circ}$C, of Schottky diodes of Layout 3 were measured at OPMD and RAL using the setup described in Section \ref{subsectioin:test setups at ral and opmd} and are shown in Figure \ref{fig:iv_schottkydiode_t3}. 
In the reverse biasing mode with the guard ring (GR) floating (FLT), the electrical breakdown occurs above 650 V, and the leakage current is less than 100 nA before breakdown. 
Grounding (GND) the guard ring further reduces the leakage by a factor of around 3. 
In the forward biasing mode, the current reaches around 1 mA before ohmic drop saturation becomes noticeable. The rectification ratio is close to $10^4$ with respect to the current at the forward bias of 3 V. 
As described in Appendix \ref{appendix:schottky barrier iv characteristics in the presence of interface states}, the ideality factors $n(V)$ and density of interface states $D_{is}(E)$ are extracted from the forward IV characteristics, and are shown in Figure \ref{fig:schottkydiode_idealityfactor_dis}.
To calculate $D_{is}(E)$, the assumed thickness $\delta$ of the interface layer was 2 nm, compatible with the expected value of the thickness of native oxide in silicon~\cite{10.1007/BF01774216}. 

\begin{figure}[htbp]
\centering
\includegraphics[width=0.4\textwidth]{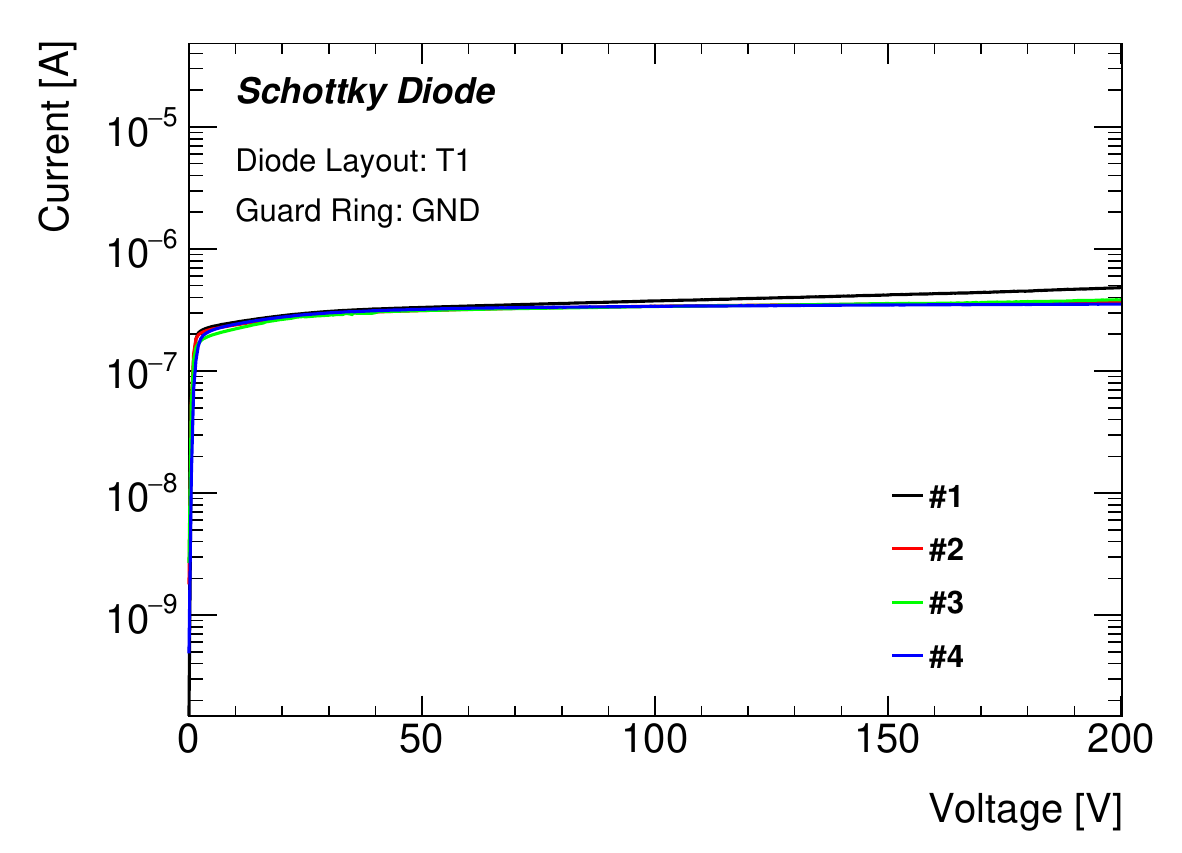}
\includegraphics[width=0.4\textwidth]{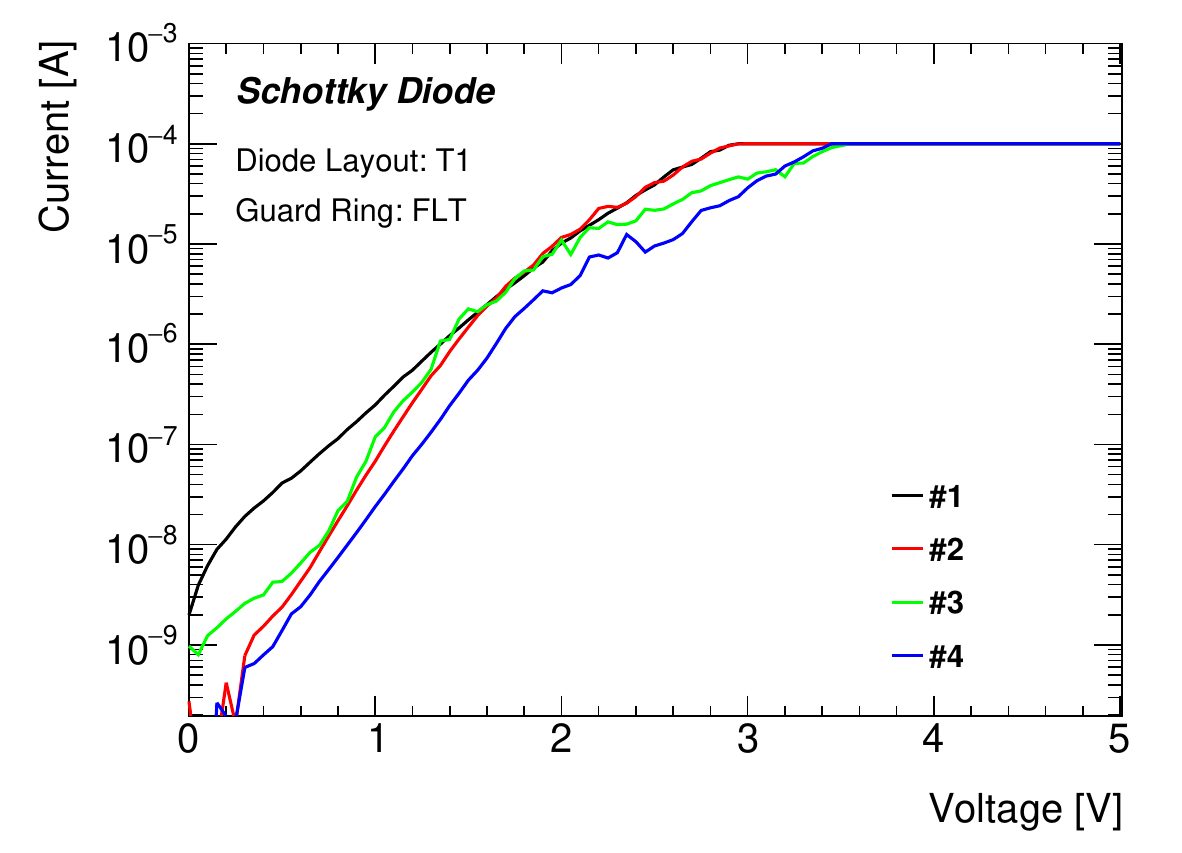}
\caption{IV characteristics of Schottky diodes of Layout 1 measured at room temperature (T = 20 $^{\circ}$C). Left: Reverse IV. Right: Forward IV.}
\label{fig:iv_schottkydiode_t1}
\end{figure}

The IV characteristics of several non-irradiated Schottky diodes of Layout 1 are shown in Figure \ref{fig:iv_schottkydiode_t1}.

\subsubsection*{\textit{pn} junction}
\label{subsubsection:pn junction-iv}

\begin{figure}[htbp]
\centering
\includegraphics[width=0.4\textwidth]{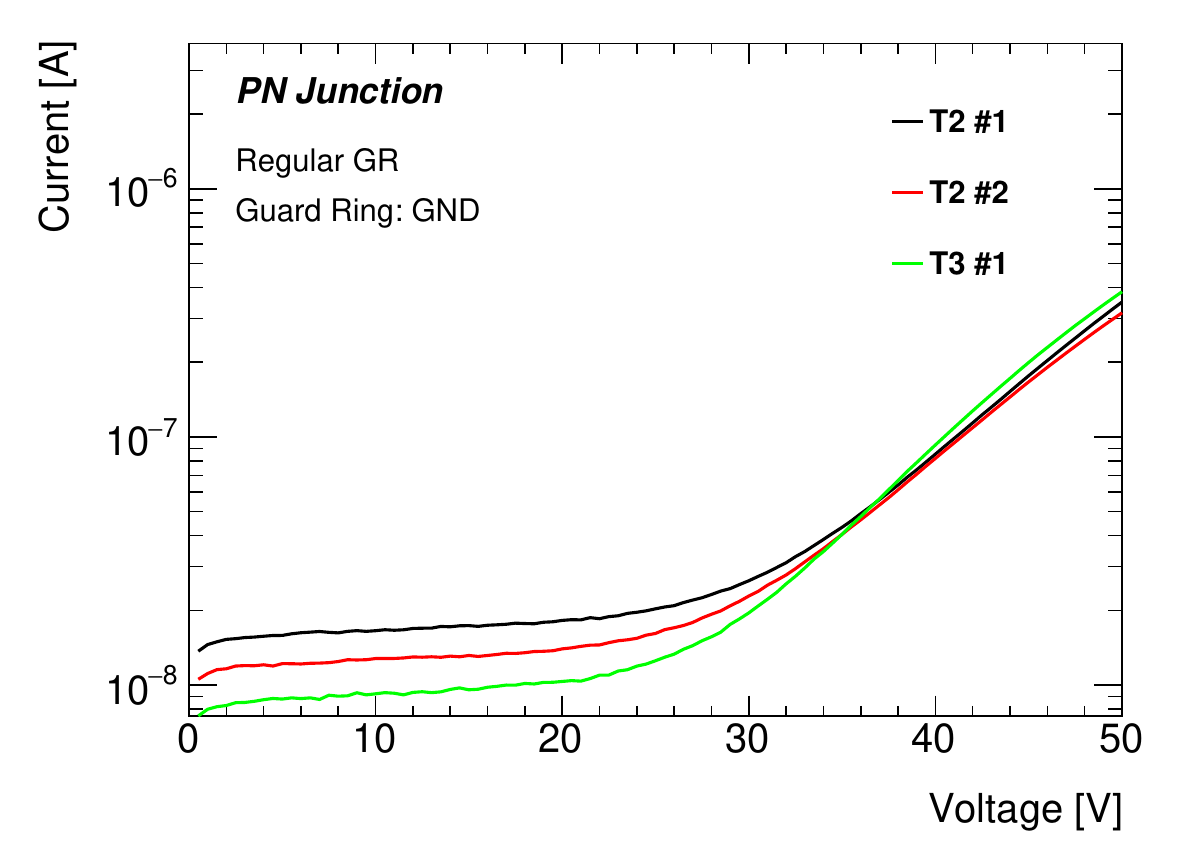}
\includegraphics[width=0.4\textwidth]{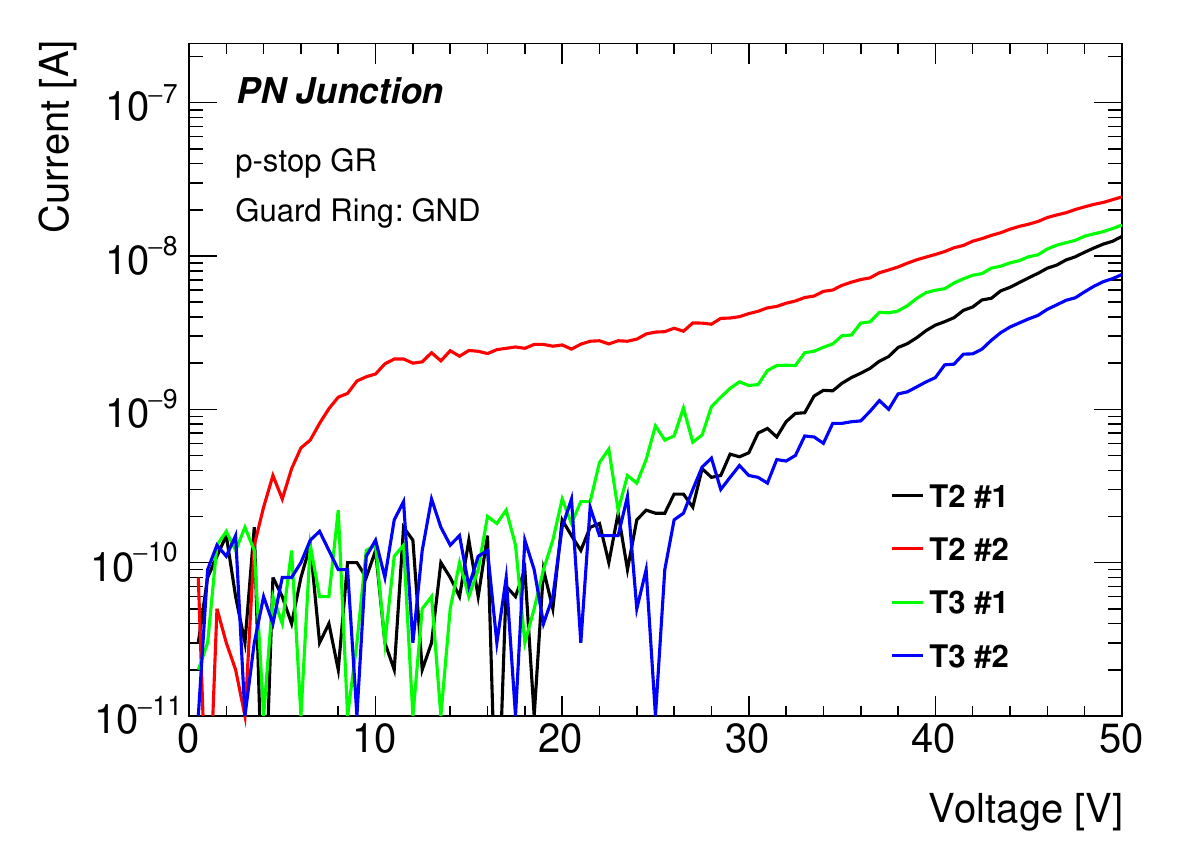}
\caption{Reverse IV characteristics of \textit{pn} junctions of Layout 2 and 3 measured at T = 22 $^{\circ}$C. Left: Regular guard ring. Right: Guard ring with \textit{p}-stop.}
\label{fig:iv_pnjunction_t2_t3}
\end{figure}

\begin{figure}[htbp]
\centering
\includegraphics[width=0.4\textwidth]{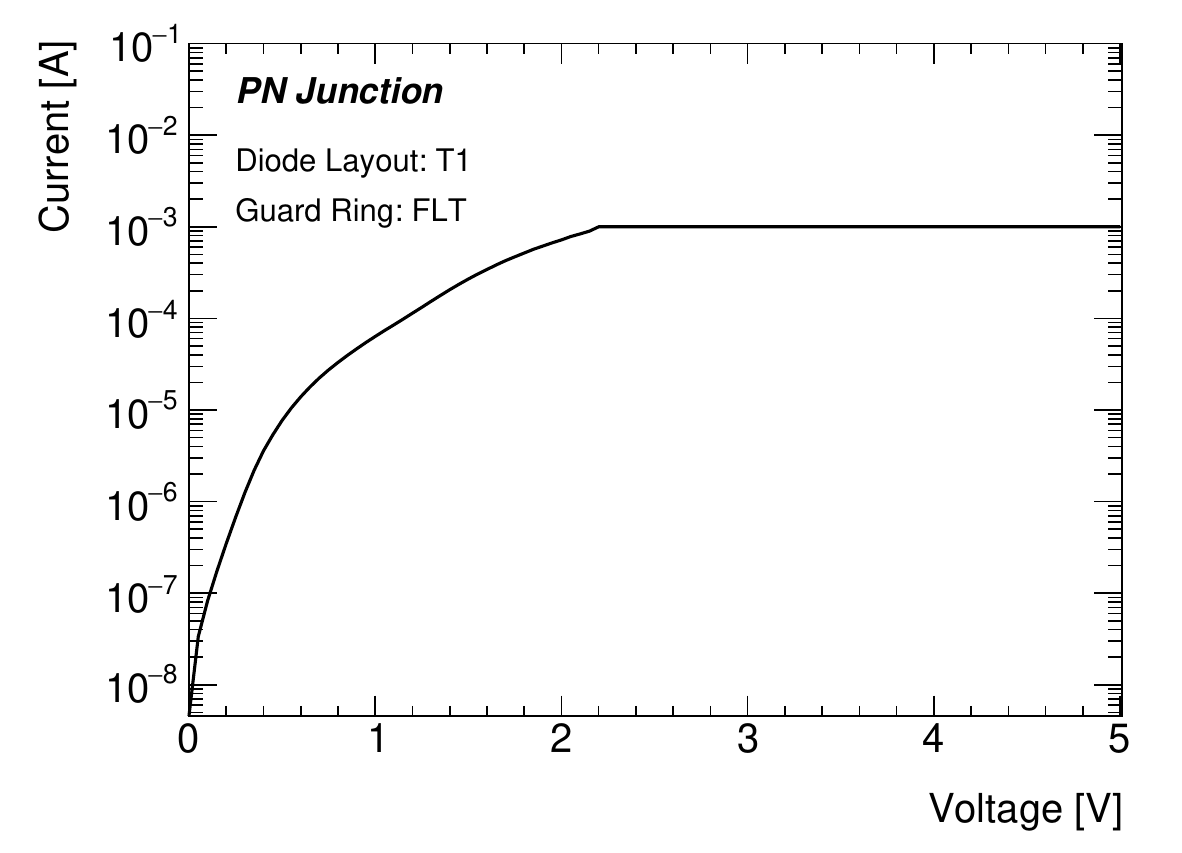}
\includegraphics[width=0.4\textwidth]{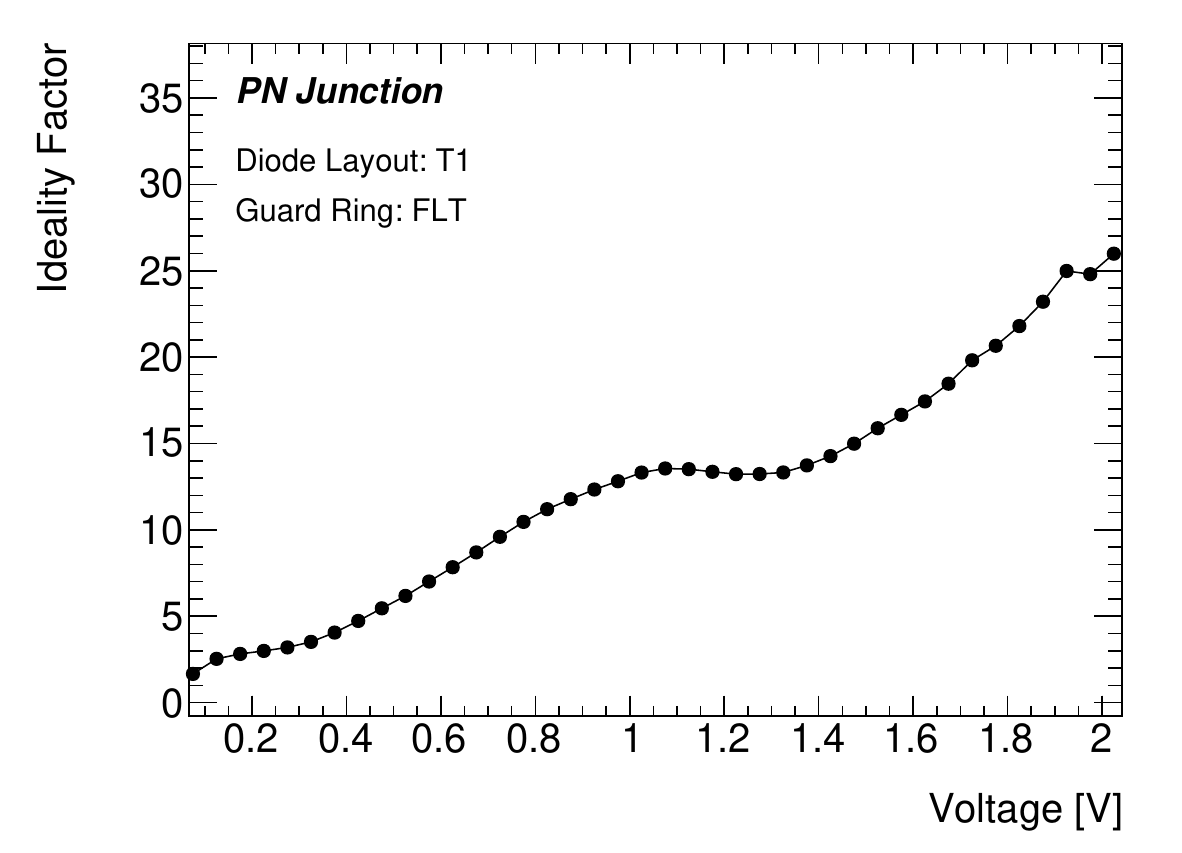}
\caption{Left: Forward IV characteristics of a \textit{pn} junction of Layout 1 at T = 22 $^{\circ}$C. Right: The ideality factor $n(V)$.}
\label{fig:iv_pnjunction_t1}
\end{figure}

The reverse IV characteristics of \textit{pn} junction devices of Layout 2 and 3 with regular guard ring (RGR) and \textit{p}-stop guard ring (PGR) are shown in Figure \ref{fig:iv_pnjunction_t2_t3}. 
The leakage current of the \textit{pn} junction of Layout 3 with regular guard ring is less than 10 nA and the leakage current of the \textit{pn} junction of Layout 2 is $\sim$15 nA up to 30 V. For the devices with \textit{p}-stop guard ring, the leakage current is hightly supressed. The cause of increase in leakage current starting at around 30 V is currently being investigated.
Figure \ref{fig:iv_pnjunction_t1} shows the forward IV characteristics and the ideality factor $n(V)$ of an example \textit{pn} junction of Layout 1.

\subsection{CV test results}
\label{subsection:cv test results}

\subsubsection*{Schottky diode}
\label{subsubsection:schottky diode-cv}

\begin{figure}[htbp]
\centering
\includegraphics[width=0.4\textwidth]{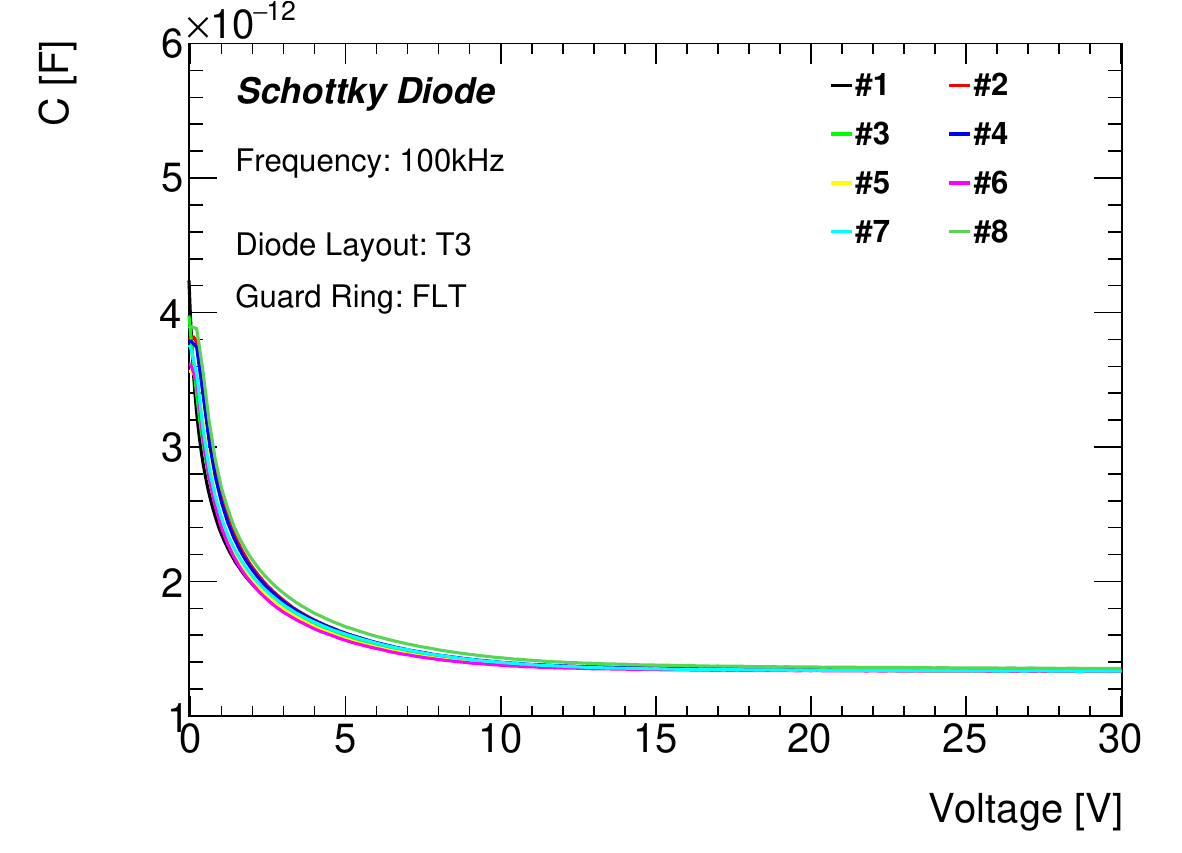}
\includegraphics[width=0.4\textwidth]{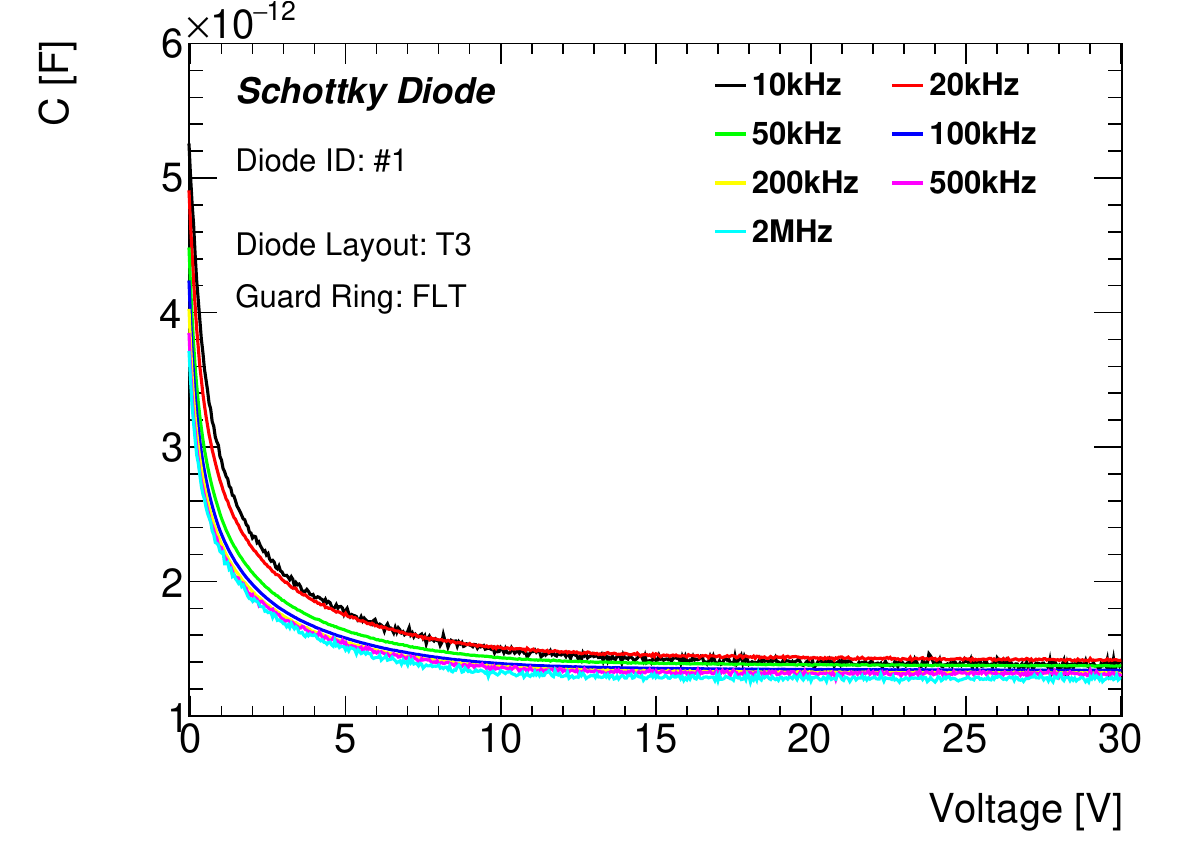}
\caption{Left: The reverse CV of Schottky diodes of Layout 3 measured at the signal frequency of 100 kHz. Right: The reverse CV versus frequency of a Schottky diodes of Layout 3.}
\label{fig:schottkydiode_cv_t3}
\end{figure}

\begin{figure}[htbp]
\centering
\includegraphics[width=0.4\textwidth]{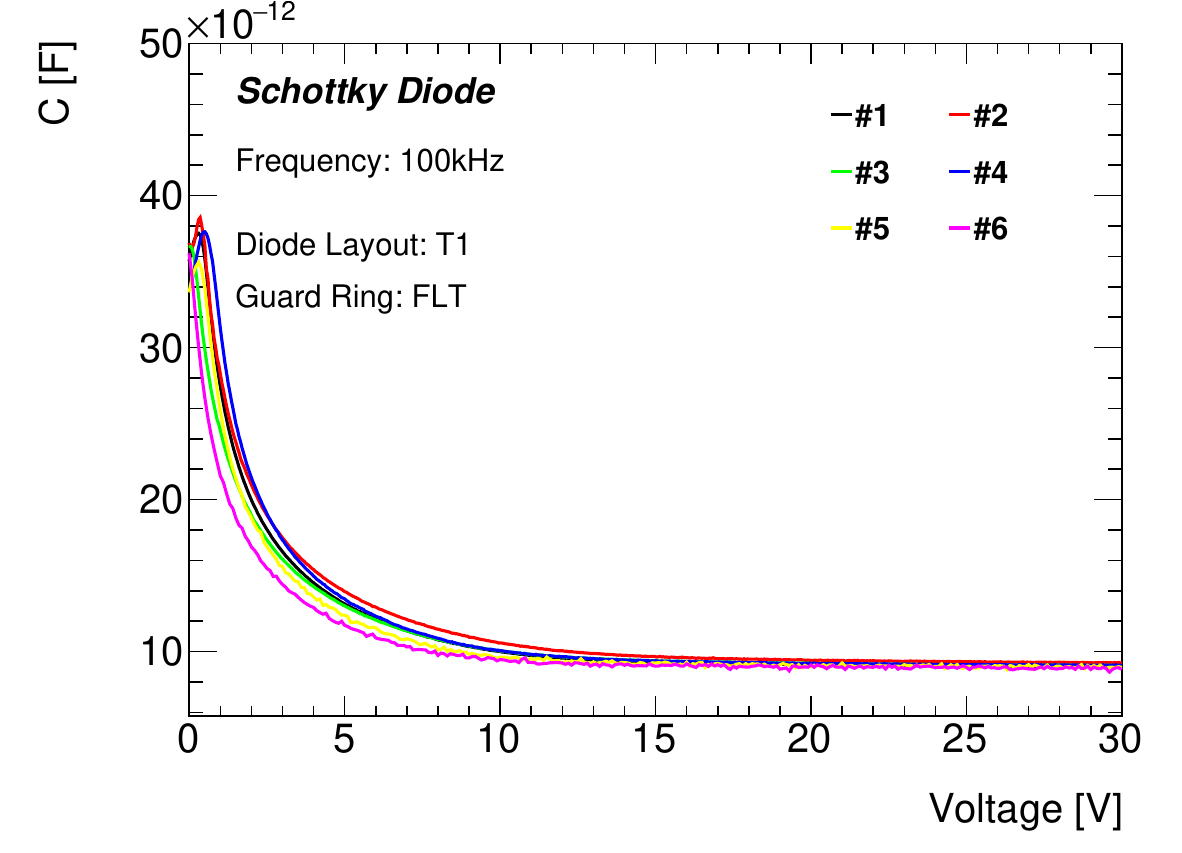}
\includegraphics[width=0.4\textwidth]{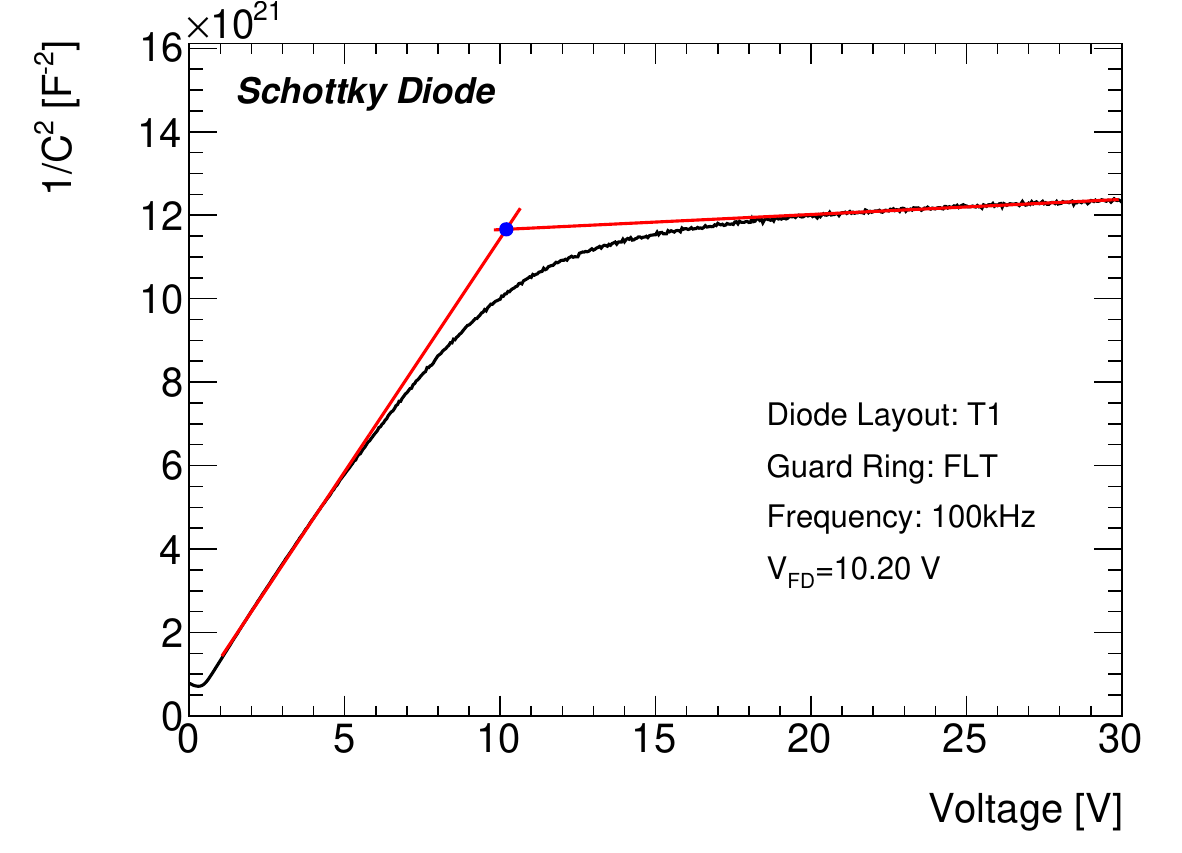}
\caption{Left: The reverse CV of Schottky diodes of Layout 1 measured at the signal frequency of 100 kHz. Right: An example of 1/C\textsuperscript{2} versus voltage of Schottky diode of Layout 1, with the extracted full depletion voltage.}
\label{fig:schottkydiode_cv_t1}
\end{figure}

In the CV measurement of Schottky diodes, the AC signal amplitude was 100 mV and its frequency ranged from 10 kHz to 2 MHz. All measuremenets were performed at 20 $^{\circ}$C. 
Figure \ref{fig:schottkydiode_cv_t3} shows the CV curves of the Schottky diode of Layout 3 under reverse biasing, the left figure shows the CV curves of 8 devices measured at the signal frequency of 100 kHz and the right figure shows the CV curves of one device measured at various frequencies. 
The cleanest CV characteristics were obtained at 100 kHz, therefore next CV plots are shown at this frequency.
Due to the smaller error introduced by the lateral extension of the depletion region beyond the cathode electrode in calculating the effective area of the capacitor, here only the devices of the Layout 1 are used to extrapolate the full depletion voltage.
Figure \ref{fig:schottkydiode_cv_t1} shows the CV curves of the Schottky diodes of Layout 1 and an example of 1/C\textsuperscript{2} versus voltage.
The fully depleted voltage is extracted using the procedure described in Appendix \ref{appendix:schottky capacitance in presence of interface states}, and it is $\sim$10 V for Schottky diode of Layout 1.

Table \ref{table:schottkydiode_parameter} shows the Schottky diode parameter values obtained from IV and CV measurements.

\begin{table}[htbp]
\centering
\caption{Parameters of Schottky diodes of Layout 1 and 3.}
\label{table:schottkydiode_parameter}
\begin{tabular}{llcc}
\toprule
Sample & Parameter & Average & $\sigma$ \\
\midrule
\multirow{4}{*}{Schottky diode T3} & Barrier height $\Phi_{Bp0}$ [V] & 0.744 & 0.014 \\
 & Ideality factor $n(0)$ & 3.72 & 0.97 \\
 & Breakdown Voltage [V] & >650/700 & -- \\
 & Leakage current density at 50 V [A$\cdot$cm\textsuperscript{-2}] & 1.35$\times$10\textsuperscript{-5} & 6.28$\times$10\textsuperscript{-7}\\
\midrule
\multirow{4}{*}{Schottky diode T1} & Barrier height $\Phi_{Bp0}$ [V] & 0.88 & 0.072 \\
 & Ideality factor $n(0)$  & 3.32 & 0.6 \\
 & Full depletion voltage $V_{FD}$ [V]  & 9.84 & 0.71 \\
 & Leakage current density at 50 V [A$\cdot$cm\textsuperscript{-2}]  & 1.01$\times$10\textsuperscript{-5} & 2.9$\times$10\textsuperscript{-7} \\
\bottomrule
\end{tabular}
\end{table}

\subsubsection*{\textit{pn} junction}
\label{subsubsection:pn junction-cv}
\begin{figure}[htbp]
\centering
\includegraphics[width=0.6\textwidth]{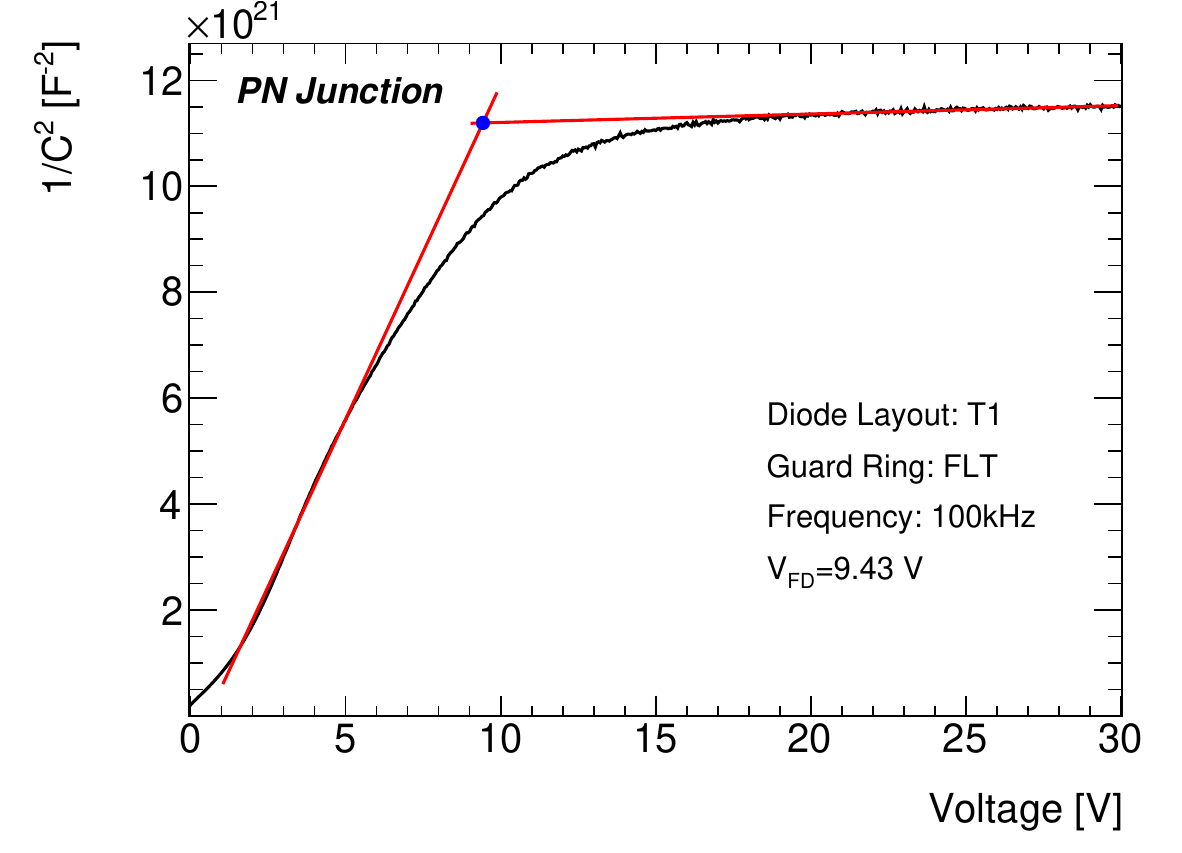}
\caption{An example of 1/C\textsuperscript{2} versus voltage of \textit{pn} junction of Layout 1 and the extrapolated full depletion voltage.}
\label{fig:pnjunction_cv_t1}
\end{figure}

The plot of Figure \ref{fig:pnjunction_cv_t1} shows the CV characteristics of one \textit{pn} junction with the regular guard ring of the Layout 1 at the signal frequency of 100 kHz. The extrapolated full depletion voltage is 9.43 V and it is consistent with the value obtained from the Schottky diode and the expected theorical value of the abrupt \textit{pn} junction.

\subsection{Charge collection efficiency test results}
\label{subsection:cce results}
The CCE of the Schottly diode of Layout 3 and \textit{pn} junction of Layout 2 were measured using the laser setup described in Section \ref{subsectioin:test setups at ral and opmd} and scanning across the central region of the device in steps of 5 $\mu$m. The CCE values shown in this paper are the integral charges of the signal pulse.

\subsubsection*{Schottky diode}
\label{subsubsection:schottky diode-cce}
\begin{figure}[htbp]
\centering
\includegraphics[width=0.4\textwidth]{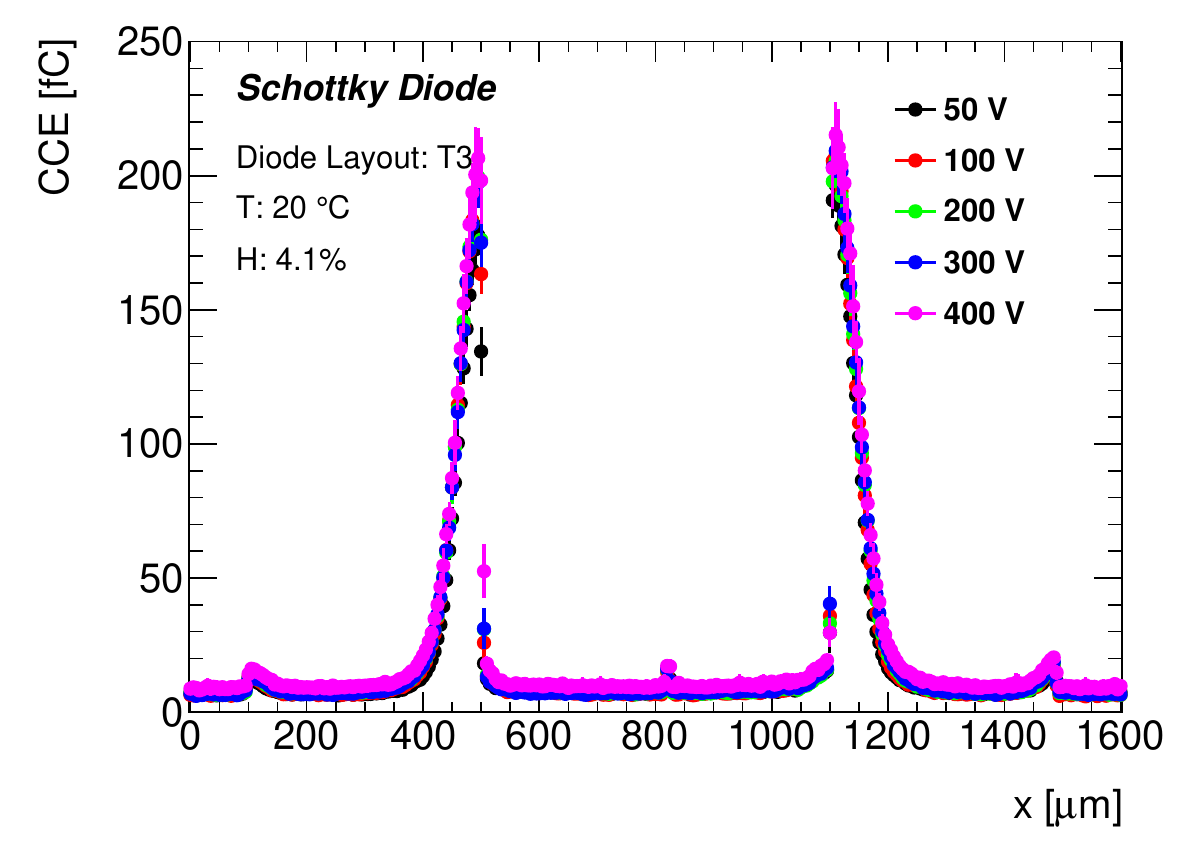}
\includegraphics[width=0.4\textwidth]{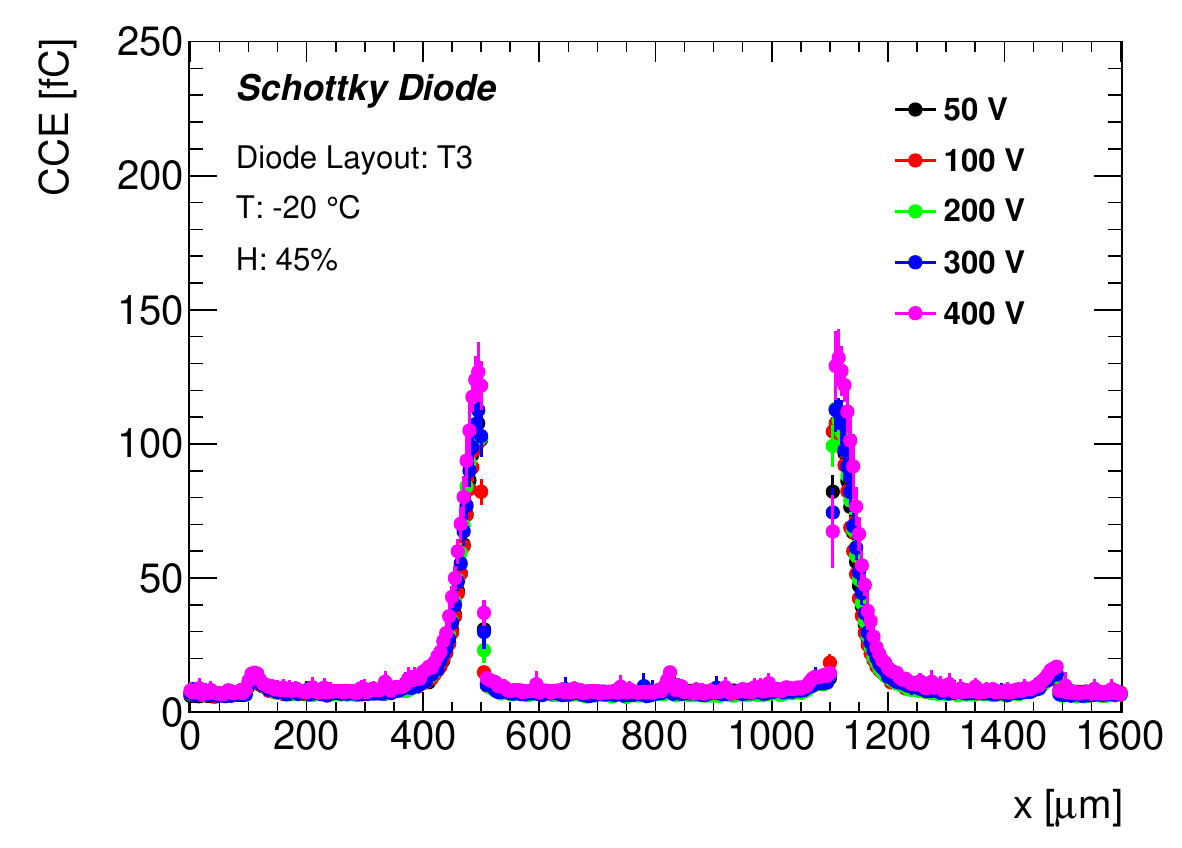}
\caption{The CCE of the Schottky diode of Layout 3 versus bias voltages at two different temperatures. Left: T = 20 $^{\circ}$C. Right: T = $-20$ $^{\circ}$C.}
\label{fig:cce_schottdiode_temperature}
\end{figure}

\begin{figure}[htbp]
\centering
\includegraphics[width=0.4\textwidth]{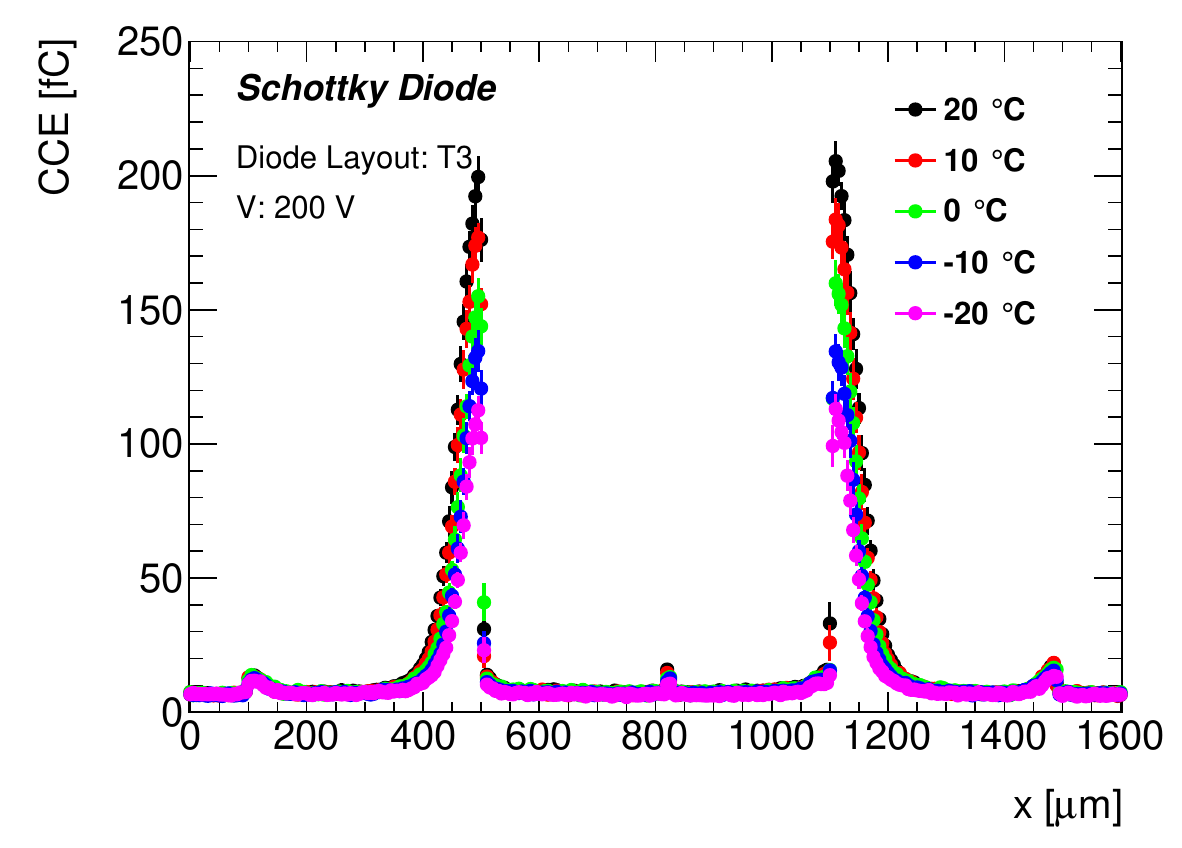}
\includegraphics[width=0.4\textwidth]{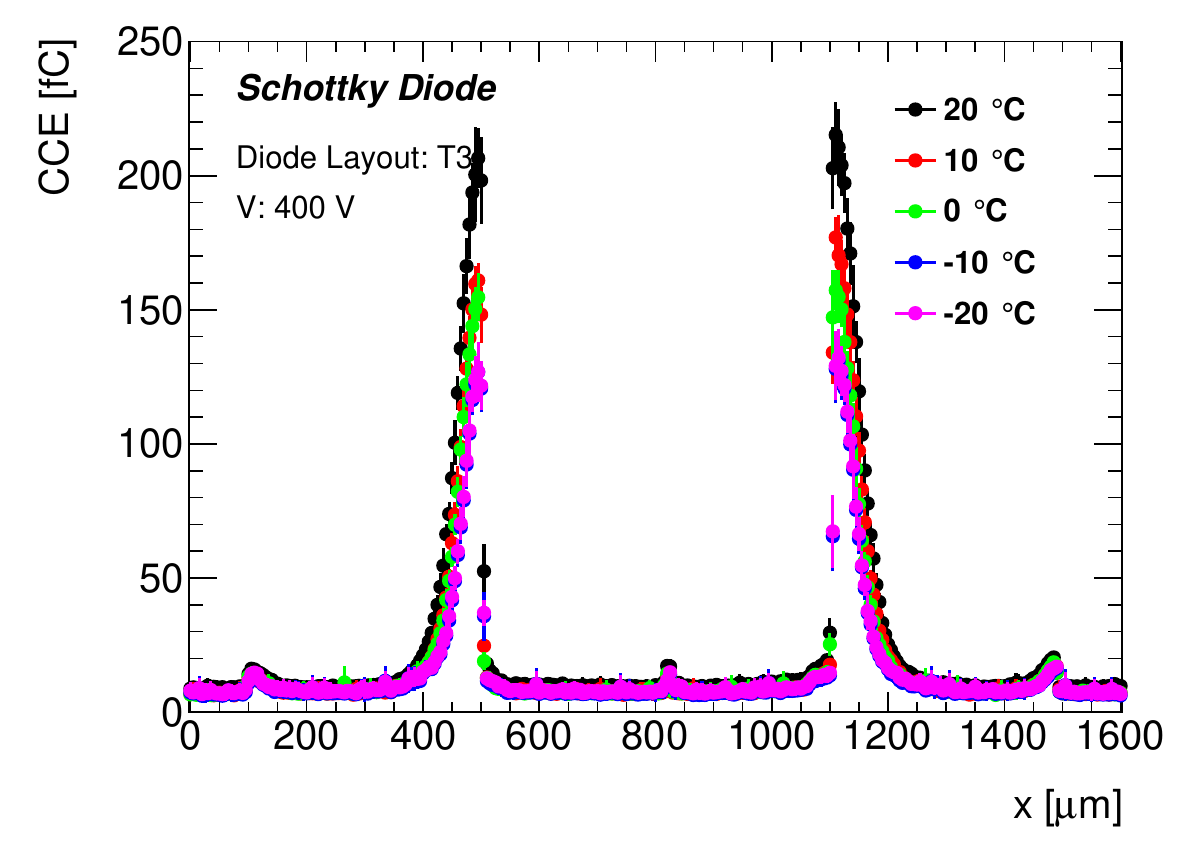}
\caption{The CCE of the Schottky diode of Layout 3 versus temepratures at two different bias voltages. Left: 200 V. Right: 400 V.}
\label{fig:cce_schottdiode_voltage}
\end{figure}

Figure \ref{fig:cce_schottdiode_temperature} and Figure \ref{fig:cce_schottdiode_voltage} show the CCE of the Schottky diode along the scanning path crossing the device center at temperatures of 20, 10, 0, $-10$, $-20$ $^{\circ}$C and bias voltages of 50, 100, 200, 300, 400 V.
The peaks in all figures represent the CCEs on the left and right edge of the cathode with the distance between the two peaks of $\sim$600 $\mu$m, corresponding to the cathode size plus the overlapping metal. 
The effect of the decrease of the IR absorption coefficient due to the temperature~\cite{https://doi.org/10.1002/pip.3474} is evident in all figures, with a significant decrease of CCE from the temperature of 20 $^{\circ}$C to $-20$ $^{\circ}$C, as shown in Figure \ref{fig:cce_schottdiode_temperature}. 
The lateral extension of the depletion region changes marginally with the increase of bias voltage as shown in Figure \ref{fig:cce_schottdiode_voltage}.

\subsubsection*{\textit{pn} junction}
\label{subsubsection:pn junction-cce}
\begin{figure}[htbp]
\centering
\includegraphics[width=0.4\textwidth]{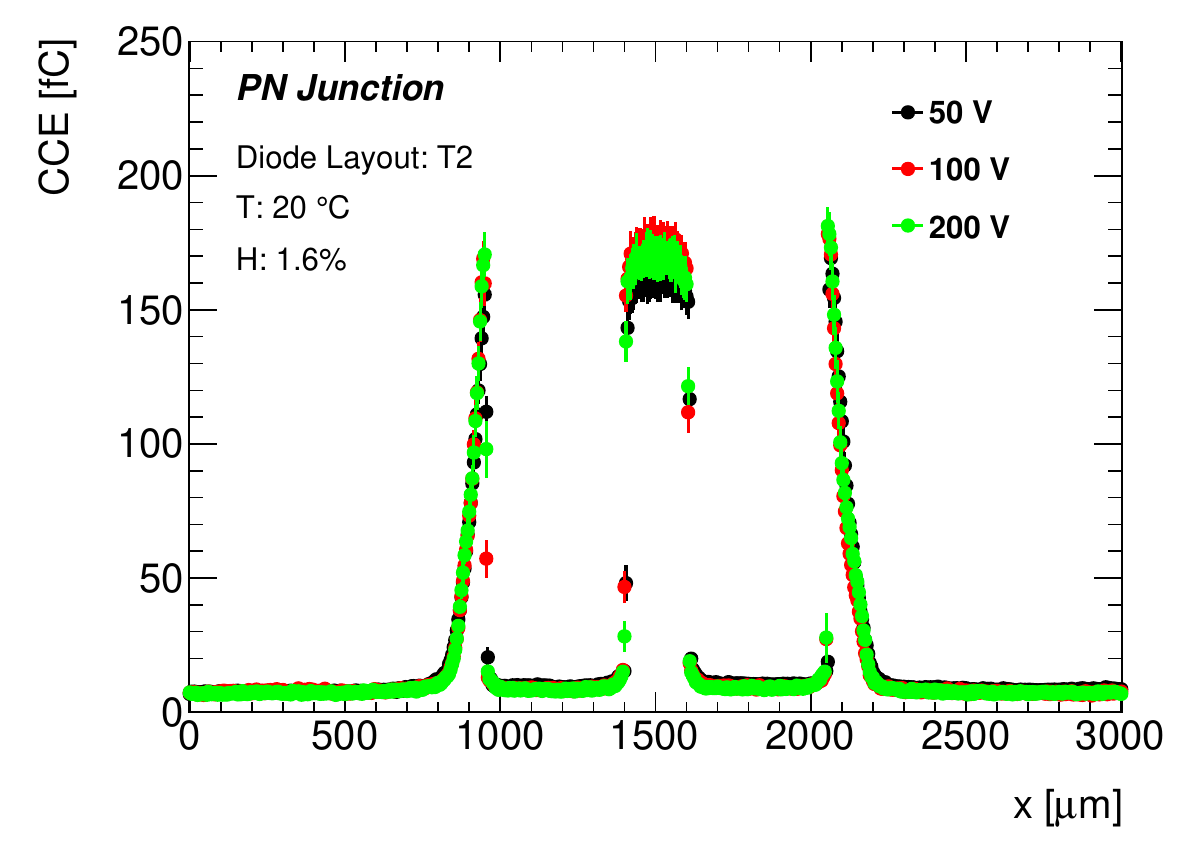}
\includegraphics[width=0.4\textwidth]{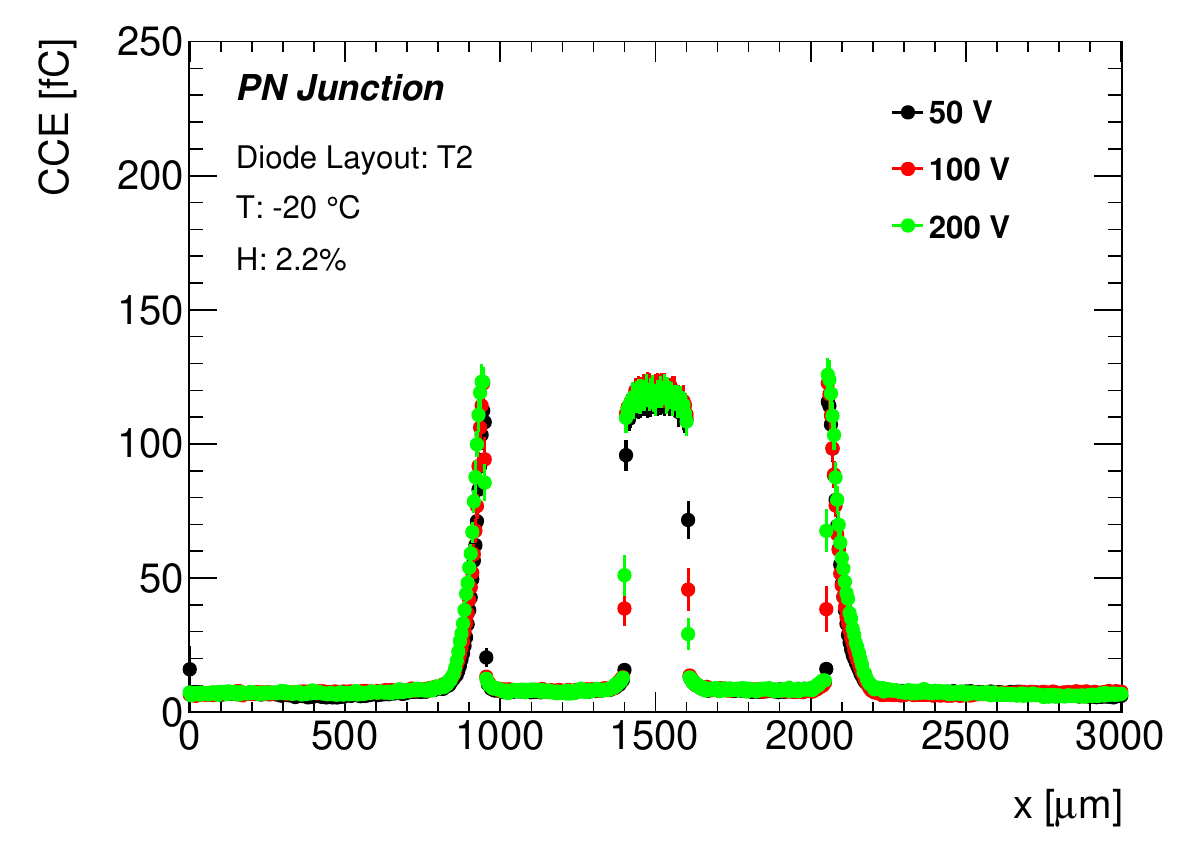}
\caption{The CCE of the \textit{pn} junction of the Layout 2 versus bias voltages at two different temperatures. Left: T = 20 $^{\circ}$C. Right: T = $-20$ $^{\circ}$C.}
\label{fig:cce_pnjunction_temperature}
\end{figure}

\begin{figure}[htbp]
\centering
\includegraphics[width=0.4\textwidth]{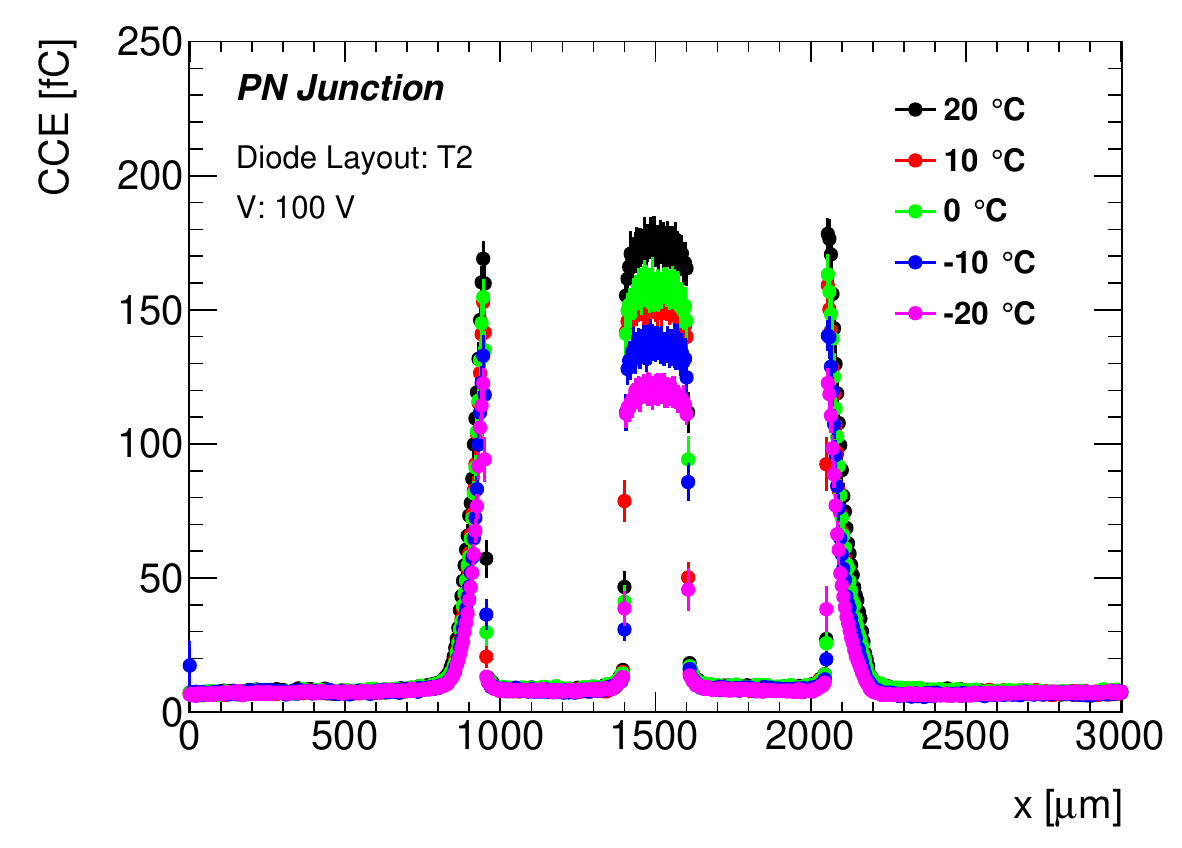}
\includegraphics[width=0.4\textwidth]{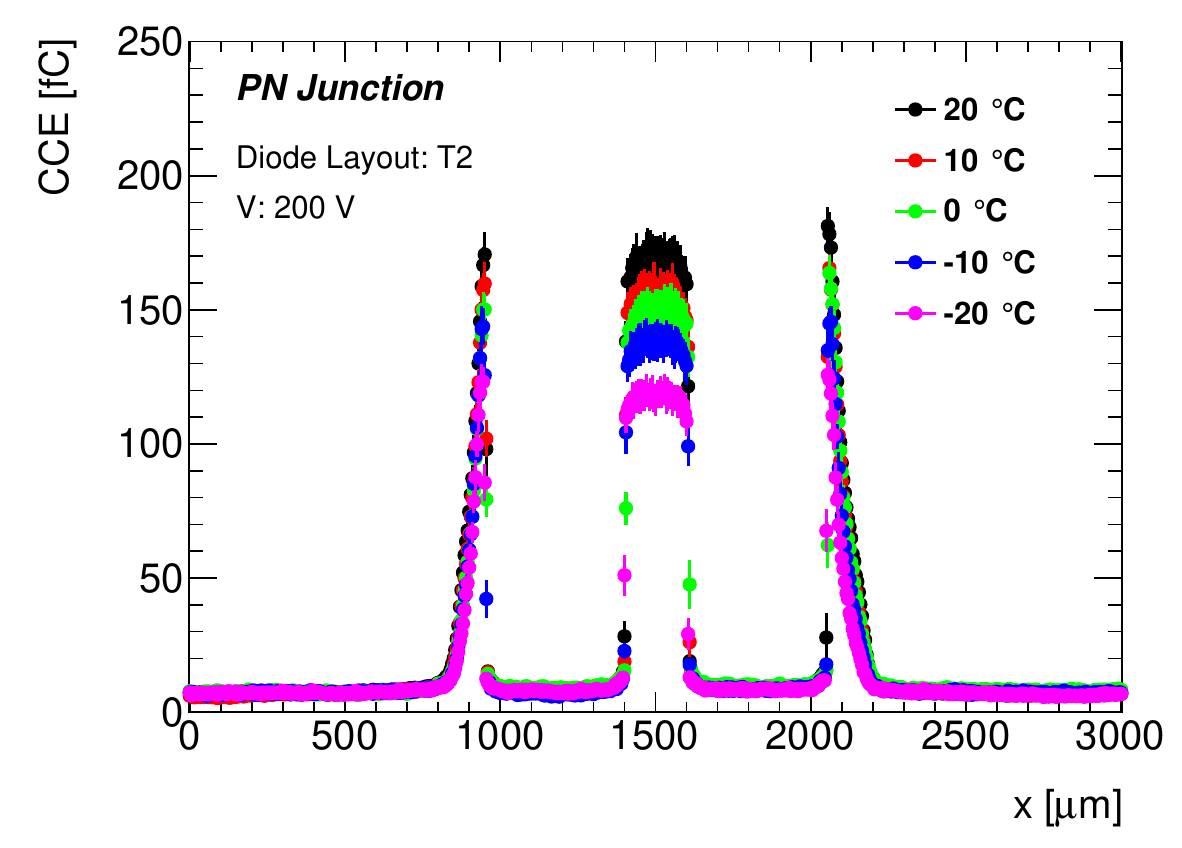}
\caption{The CCE of the \textit{pn} junction of Layout 2 versus temepratures at two different bias voltages. Left: 100 V. Right: 200 V.}
\label{fig:cce_pnjunction_voltage}
\end{figure}

Figure \ref{fig:cce_pnjunction_temperature} and Figure \ref{fig:cce_pnjunction_voltage} show the CCEs of the \textit{pn} junction of the Layout 2 along the scanning path crossing the device center at temperatures of 20, 10, 0, $-10$, $-20$ $^{\circ}$C and bias voltages of 50, 100, 200 V.
The central plateau corresponds to the central aperture of the Layout 2 with the size of $\sim$200 $\mu$m, but otherwise the CCE plot shows the same characteristics as the Schottky diode.

\subsection{DLTS results}
\label{subsection:dlts results}

The defect parameters of the unirradiated Schottky diodes and \textit{pn} junctions of various flavors and production iterations have been characterised using the DLTS setup described in Section \ref{subsection:test setups at carleton}.
These parameters will then be implemented in TCAD models to compare simulations with test results. The same procedure will be repeated in the future with irradiated samples.
For the DLTS measurements, multiple complete scans were performed for all diode samples at different bias voltages and filling pulse settings. 

\subsubsection*{\textit{pn} junction}
\label{subsubsection:pn junction-dlts}

\begin{figure}[htbp]
\centering
\includegraphics[width=0.4\textwidth]{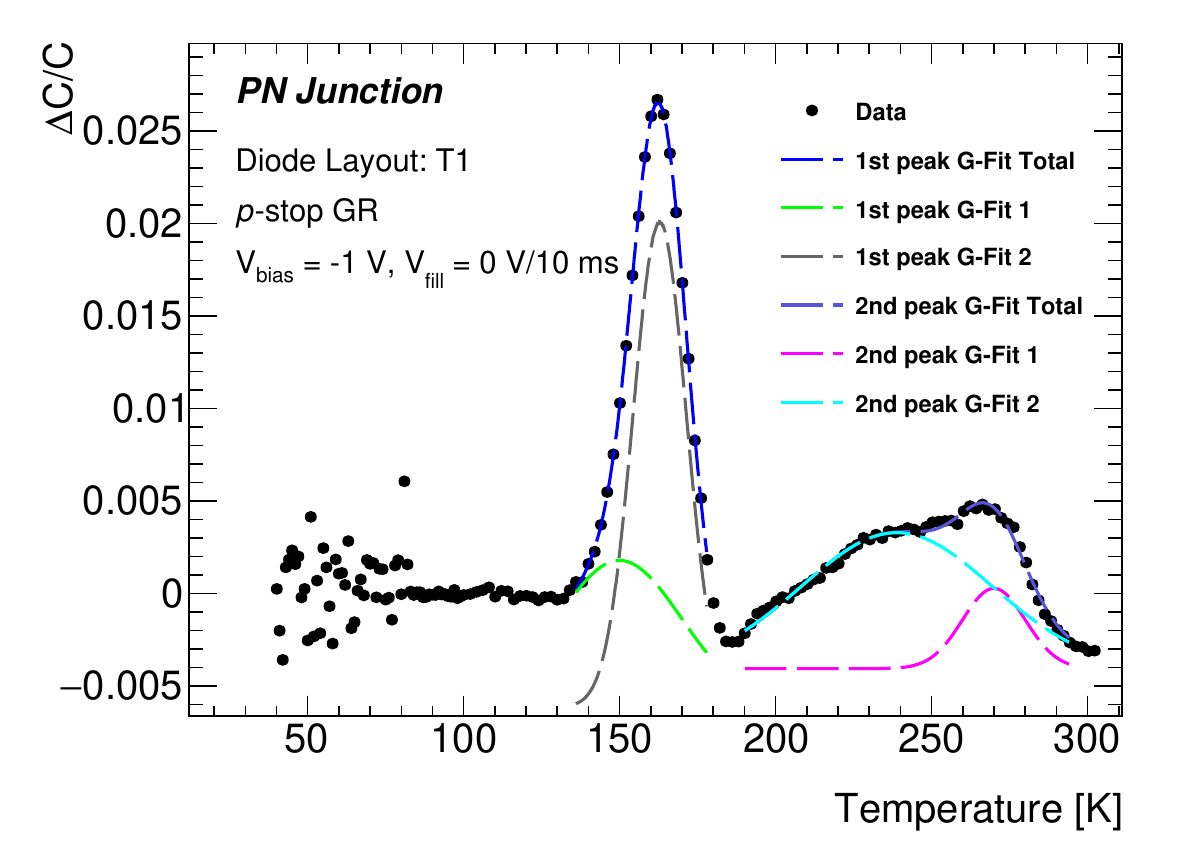}
\includegraphics[width=0.4\textwidth]{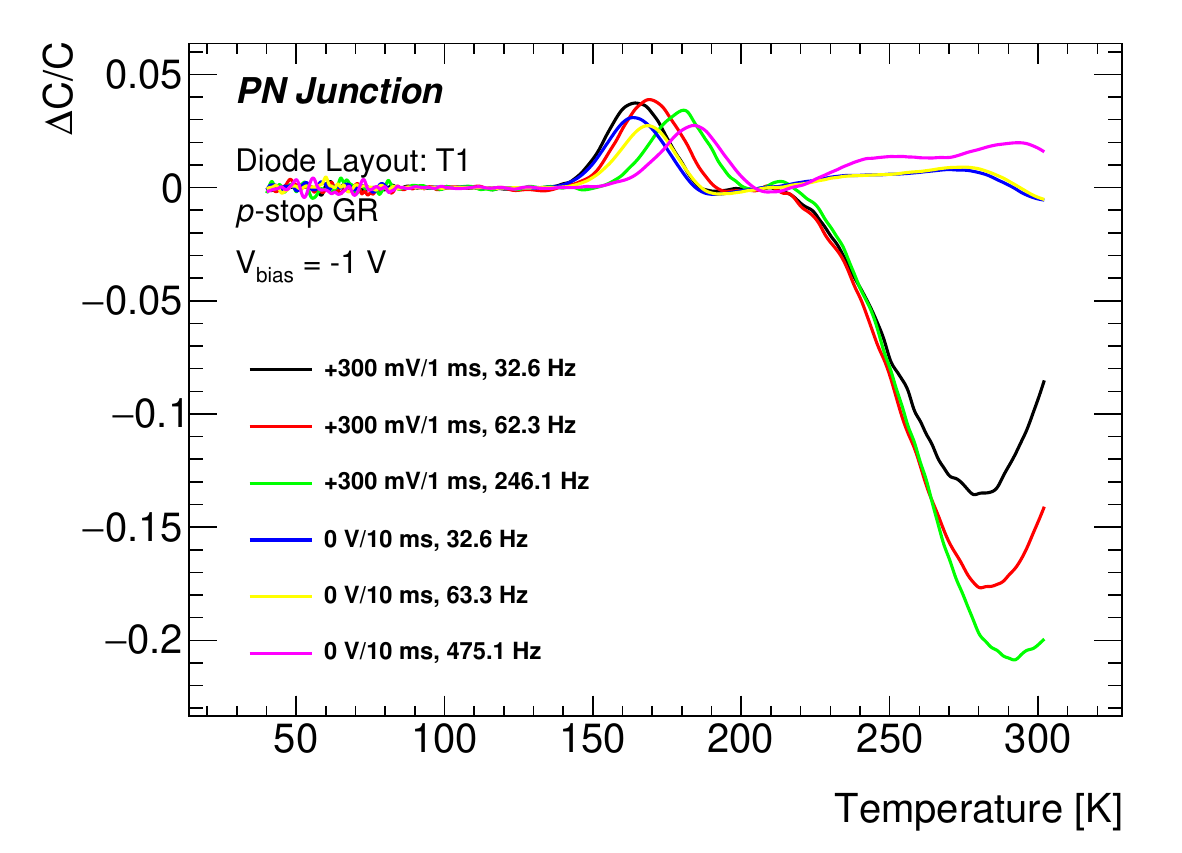}
\caption{DLTS spectra of a \textit{pn} junction with \textit{p}-stop guard ring for selected rate windows. The double-Gaussian fits to derive the peak parameters are shown on the left as an example for the general analysis procedure.}
\label{fig:dlts_pnjunction_pstopgr}
\end{figure}

\begin{figure}[htbp]
\centering
\includegraphics[width=0.3\textwidth]{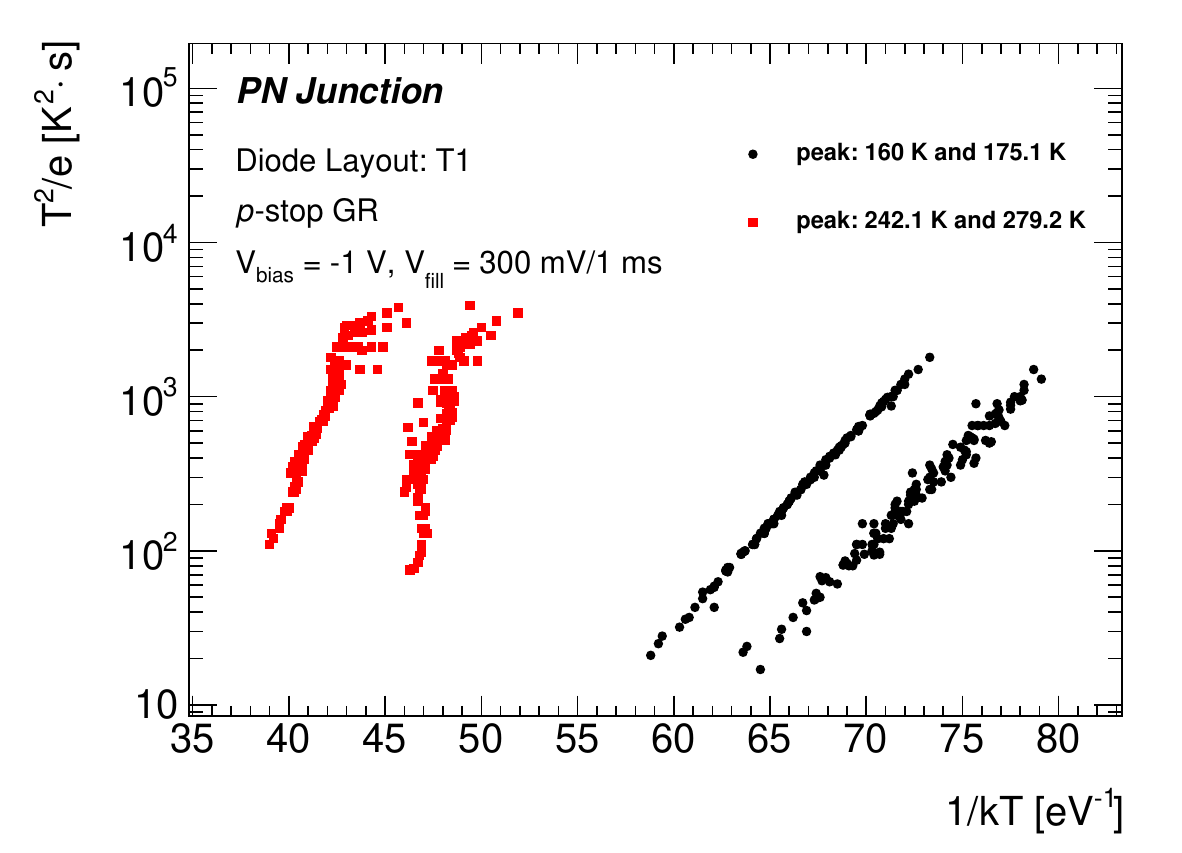}
\includegraphics[width=0.3\textwidth]{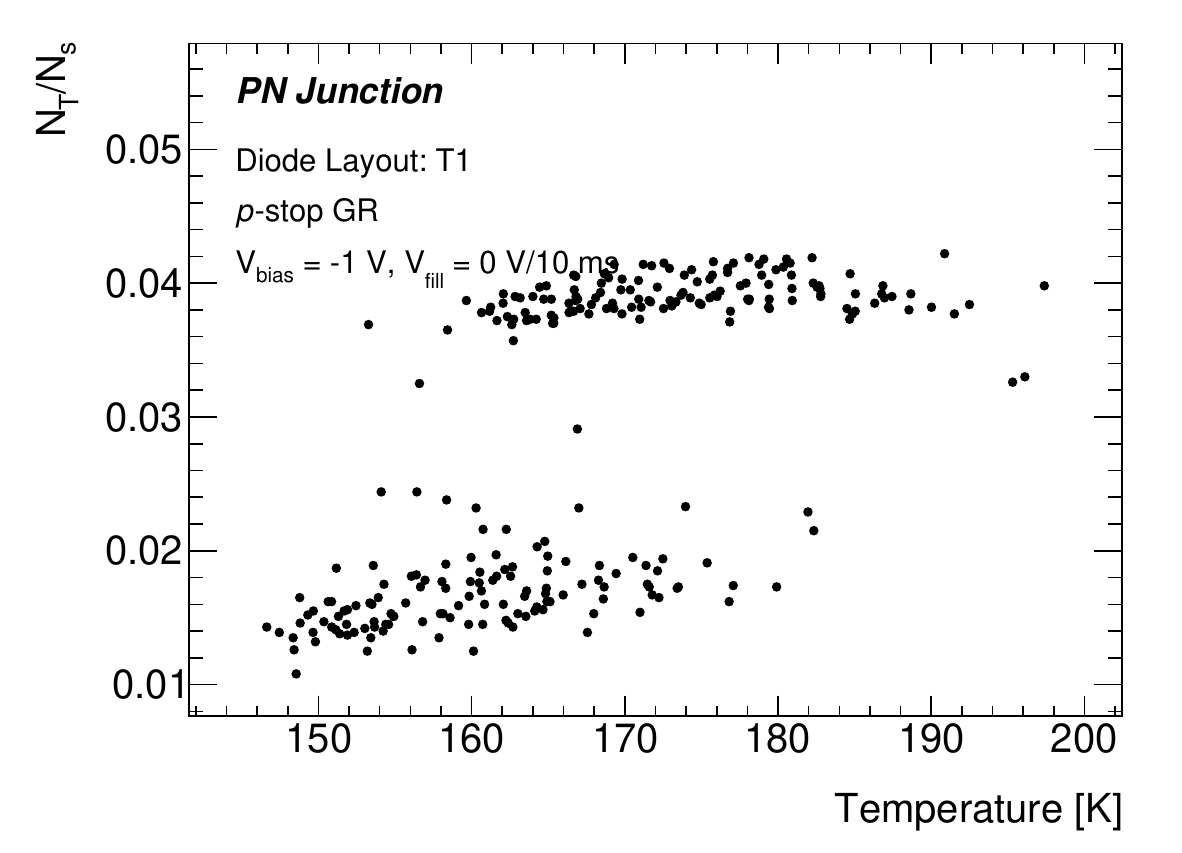}
\includegraphics[width=0.3\textwidth]{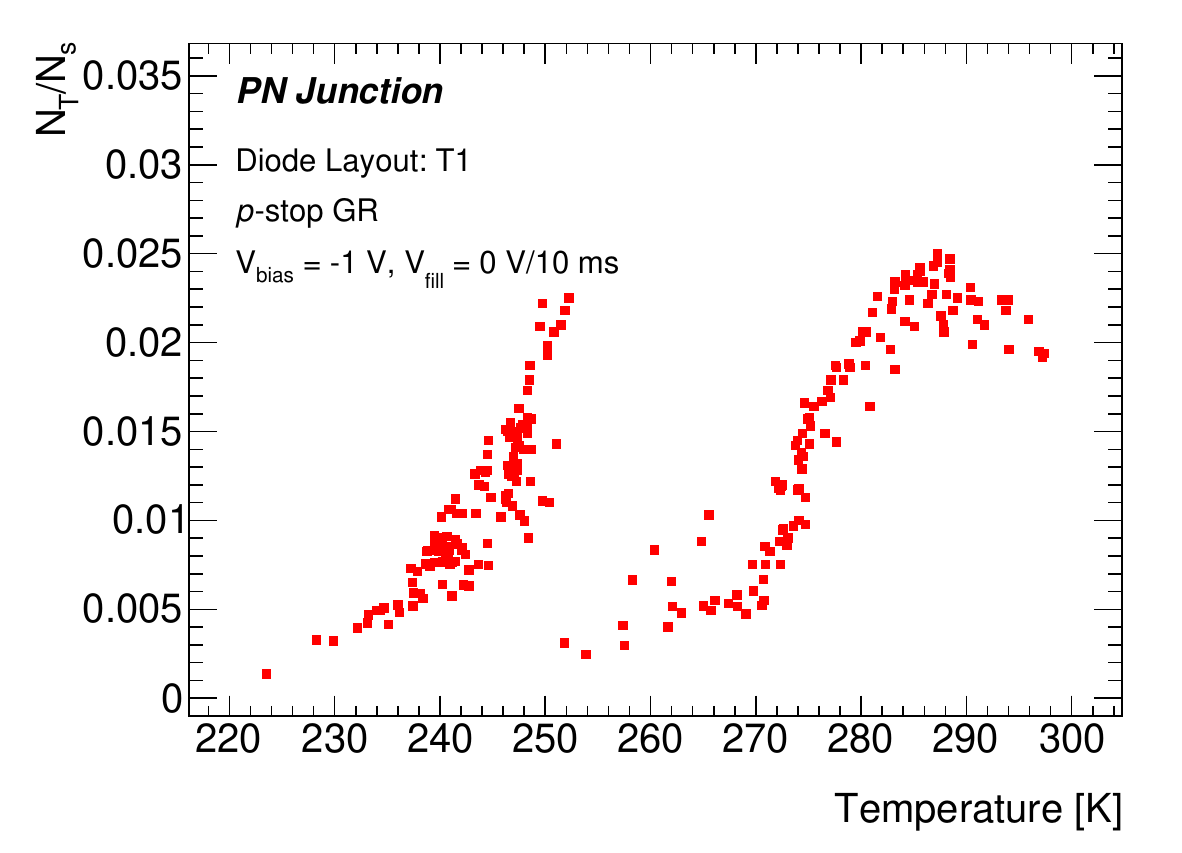}
\includegraphics[width=0.3\textwidth]{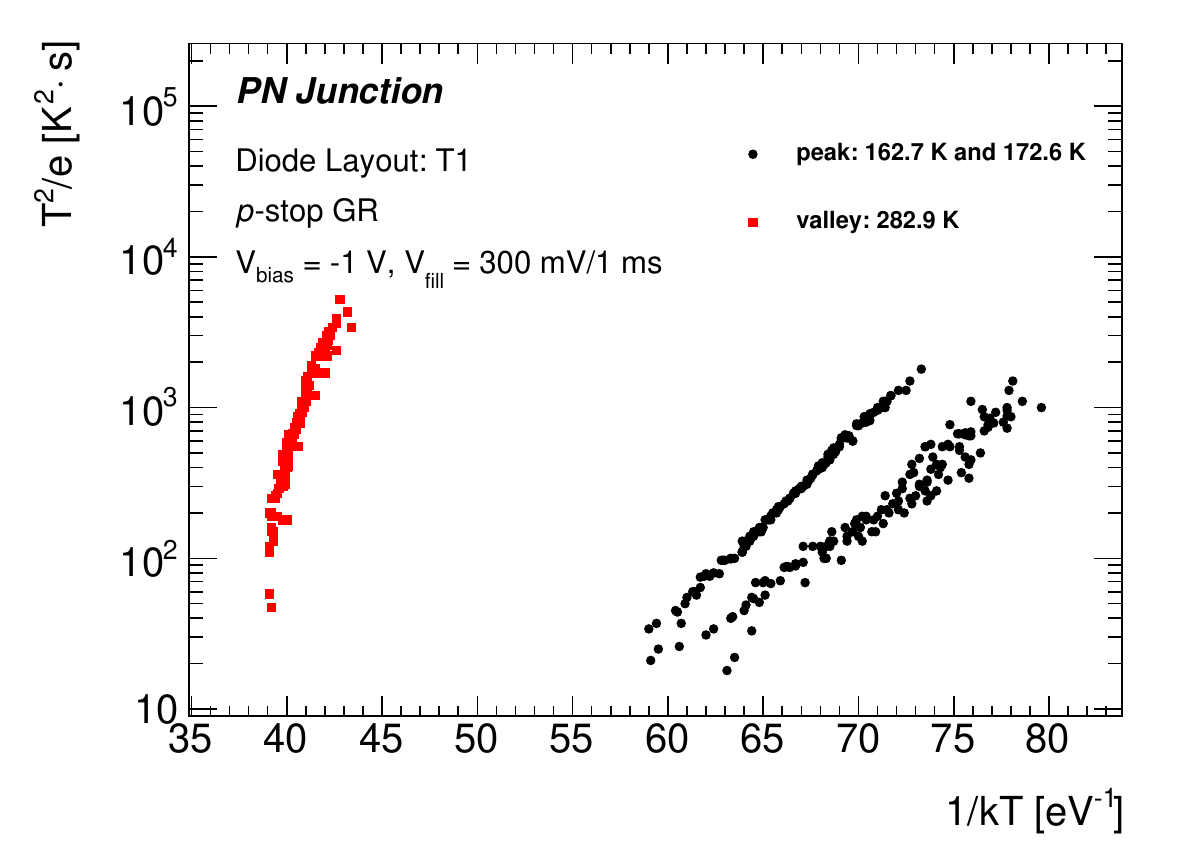}
\includegraphics[width=0.3\textwidth]{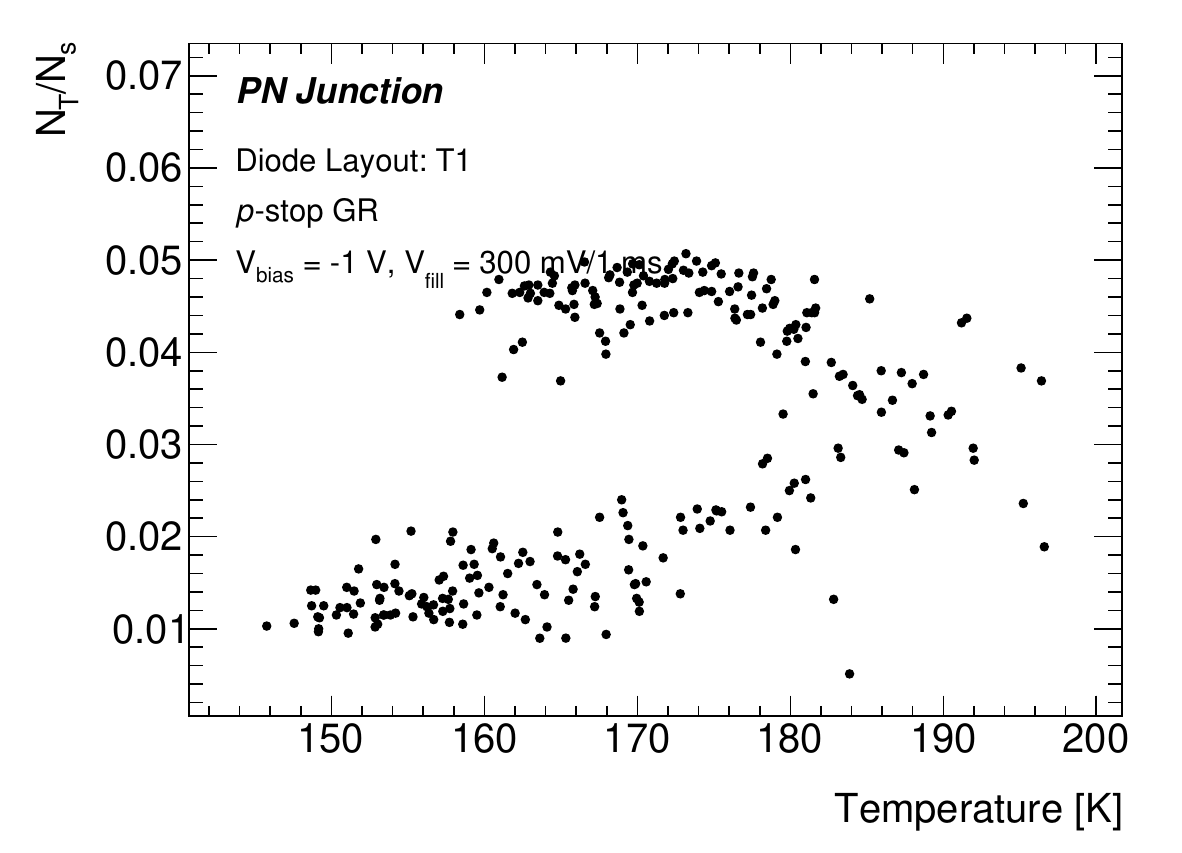}
\includegraphics[width=0.3\textwidth]{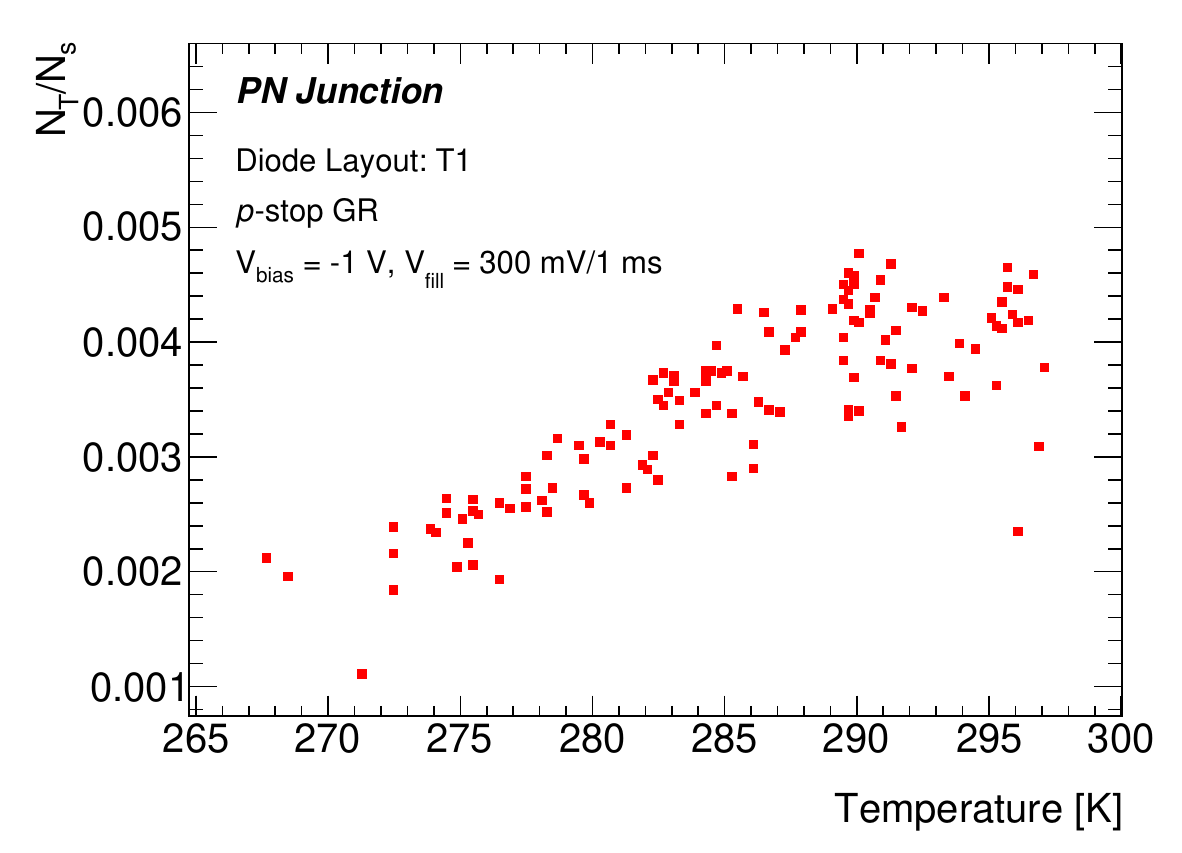}
\caption{Left: Arrhenius plots of a \textit{pn} junction with \textit{p}-stop guard ring as examples for results obtained with both hole and high injection, using reverse and forward bias pulses, respectively. Middle and Right: relative trap saturation corresponding to the data points on the Arrhenius plots.}
\label{fig:dlts_pnjunction_pstopgr_arrhenius}
\end{figure}

The left plot of Figure \ref{fig:dlts_pnjunction_pstopgr} shows the obtained DLTS spectra of a \textit{pn} junction with \textit{p}-stop guard ring. Its extrema are extracted by the double or triple-Gaussian fit algorithms.
The right plot of Figure \ref{fig:dlts_pnjunction_pstopgr} shows the comparison between DLTS spectra derived from different selected rate windows and obtained for two distinct filling pulses.
For the device with \textit{p}-stop guard ring, using a filling pulse with forward bias allows measuring minority carriers (electrons), while majority carriers (holes) are measured using reverse bias filling pulses.
The DLTS spectra did not qualitatively change with the applied bias voltage and only the observed trap saturation was affected.
DLTS spectra are used to yield the Arrhenius plots for each of the scans performed, two of which are shown in Figure \ref{fig:dlts_pnjunction_pstopgr_arrhenius} as examples. 
From both the DLTS spectra and the Arrhenius plots a good agreement is observed for the parameter causing the peak centered around 170 K.

\begin{figure}[htbp]
\centering
\includegraphics[width=0.4\textwidth]{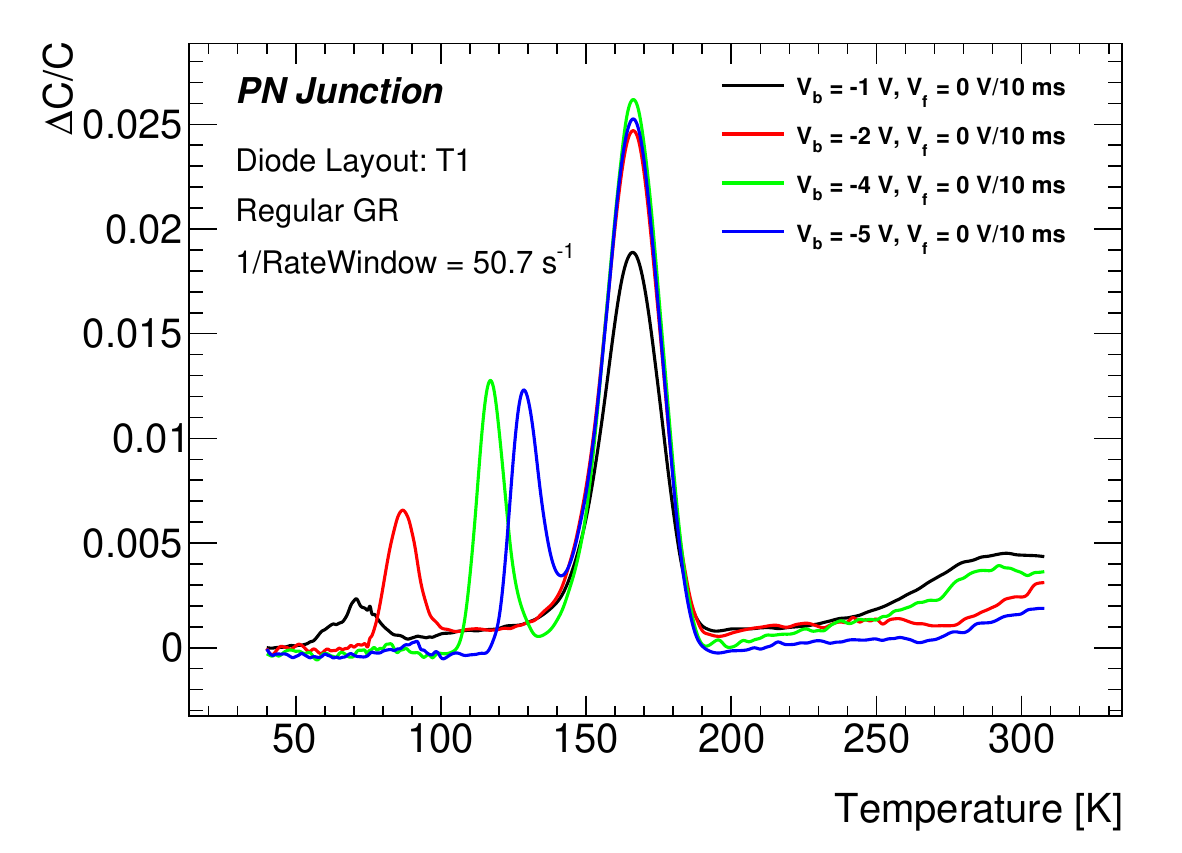}
\includegraphics[width=0.4\textwidth]{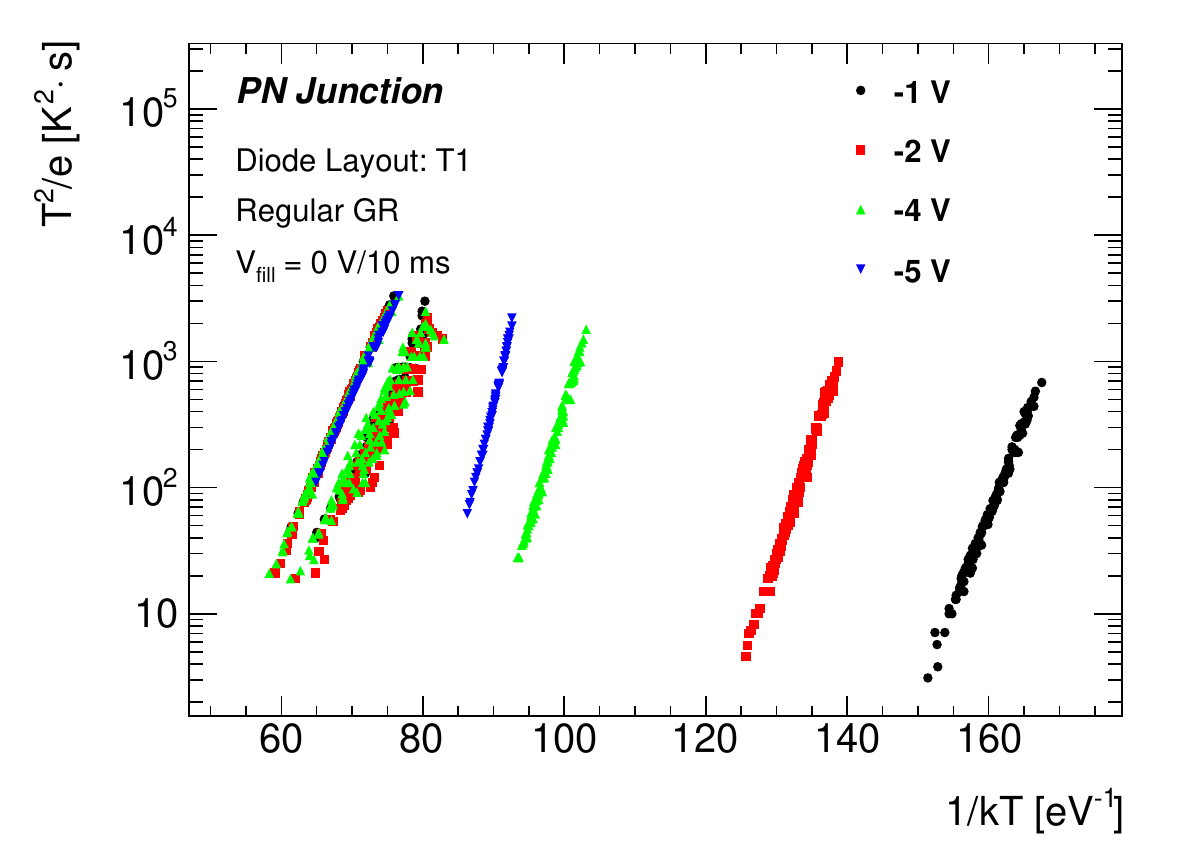}
\caption{DLTS spectra of a \textit{pn} junction without \textit{p}-stop guard ring for the same filling pulse but different applied bias voltages (left) and resulting Arrhenius plots (right).}
\label{fig:dlts_pnjunction_regulargr_arrhenius}
\end{figure}

The \textit{pn} junction samples with the regular guard ring were measured and analyzed in the same way, the DLTS spectra and the Arrhenius plots are shown in Figure \ref{fig:dlts_pnjunction_regulargr_arrhenius}. 
Unlike the samples with the \textit{p}-stop guard ring, the spectra did not vary much by changing the filling pulse voltage and there was no electron trap to be observed. 
However, the results for different bias voltages differed by a large margin and a specific peak at low temperatures showed a large dependence on the applied bias. 
This is an indication that this peak stems from surface or interface traps. 
Aside from that, the same common peak around 170 K could be observed and parameters showed good agreement between all scans. 
An additional onset of a peak at higher-than-room-temperature can be seen in the spectrum, but an analysis was not possible due to the limited temperature range of the scan.

Table \ref{table:dlts_pnjunction} shows the trap parameters obtained from the DLTS measurements.

\begin{table}[htbp]
\centering
\caption{Trap parameters obtained from DLTS measurements for \textit{pn} junction diode samples. Common peaks between different scans are represented by a range of observed parameters.}
\label{table:dlts_pnjunction}
\begin{tabular}{llccc}
\toprule
Sample & Scan & T\textsubscript{median} [K] & E\textsubscript{trap} [eV] & $\sigma$ [cm\textsuperscript{2}] \\
\midrule
\multirow{5}{*}{\textit{pn} junction \textit{p}-stop GR} &  V\textsubscript{fill} = 0 V/10 ms & 242.7 & N/A & N/A \\
 & V\textsubscript{fill} = 0 V/10 ms & 276.2 & 0.59$\pm$0.04 & 2.4$\times$10\textsuperscript{-14}$\pm$5.4$\times$10\textsuperscript{-14}  \\
 & V\textsubscript{fill} = 0 V/10 ms & 276.2 & 0.59$\pm$0.04 & 2.4$\times$10\textsuperscript{-14}$\pm$5.4$\times$10\textsuperscript{-14}  \\
 & V\textsubscript{fill} = 300 mV/1 ms & 282.9 & -0.832$\pm$0.037 & 1.8$\times$10\textsuperscript{-10}$\pm$4.5$\times$10\textsuperscript{-10}  \\ \cline{2-5}
 & \multirow{2}{*}{common peaks} & 160$-$163 & 0.217$-$0.268 & 7.8$\times$10\textsuperscript{-18}-3.0$\times$10\textsuperscript{-16}  \\
 &  & 172$-$175 & 0.287$-$0.309 & 2.2$\times$10\textsuperscript{-16}-1.0$\times$10\textsuperscript{-15}  \\
\midrule

\multirow{6}{*}{\textit{pn} junction regular GR} &  V\textsubscript{bias} = $-$1 V & 72.6 & 0.330$\pm$0.007 & 4.1$\times$10\textsuperscript{-1}$\pm$3.1$\times$10\textsuperscript{-1} \\
 &  V\textsubscript{bias} = $-$2 V & 87.4 & 0.407$\pm$0.005 & 9.4$\times$10\textsuperscript{-1}$\pm$2.0$\times$10\textsuperscript{-1} \\
 &  V\textsubscript{bias} = $-$4 V & 118.6 & 0.442$\pm$0.005 & 9.9$\times$10\textsuperscript{-6}$\pm$1.6$\times$10\textsuperscript{-6} \\
 &  V\textsubscript{bias} = $-$5 V & 129.2 & 0.545$\pm$0.007 & 1.0$\times$10\textsuperscript{-3}$\pm$1.9$\times$10\textsuperscript{-3} \\ \cline{2-5}
 & \multirow{2}{*}{common peaks} & 157$-$159 & 0.241$-$0.260 & 4.7$\times$10\textsuperscript{-17}-1.9$\times$10\textsuperscript{-16}  \\
 &  & 169$-$172 & 0.296$-$0.303 & 4.9$\times$10\textsuperscript{-16}-8.3$\times$10\textsuperscript{-15}  \\
\bottomrule
\end{tabular}
\end{table}

\begin{figure}[htbp]
\centering
\includegraphics[width=0.6\textwidth]{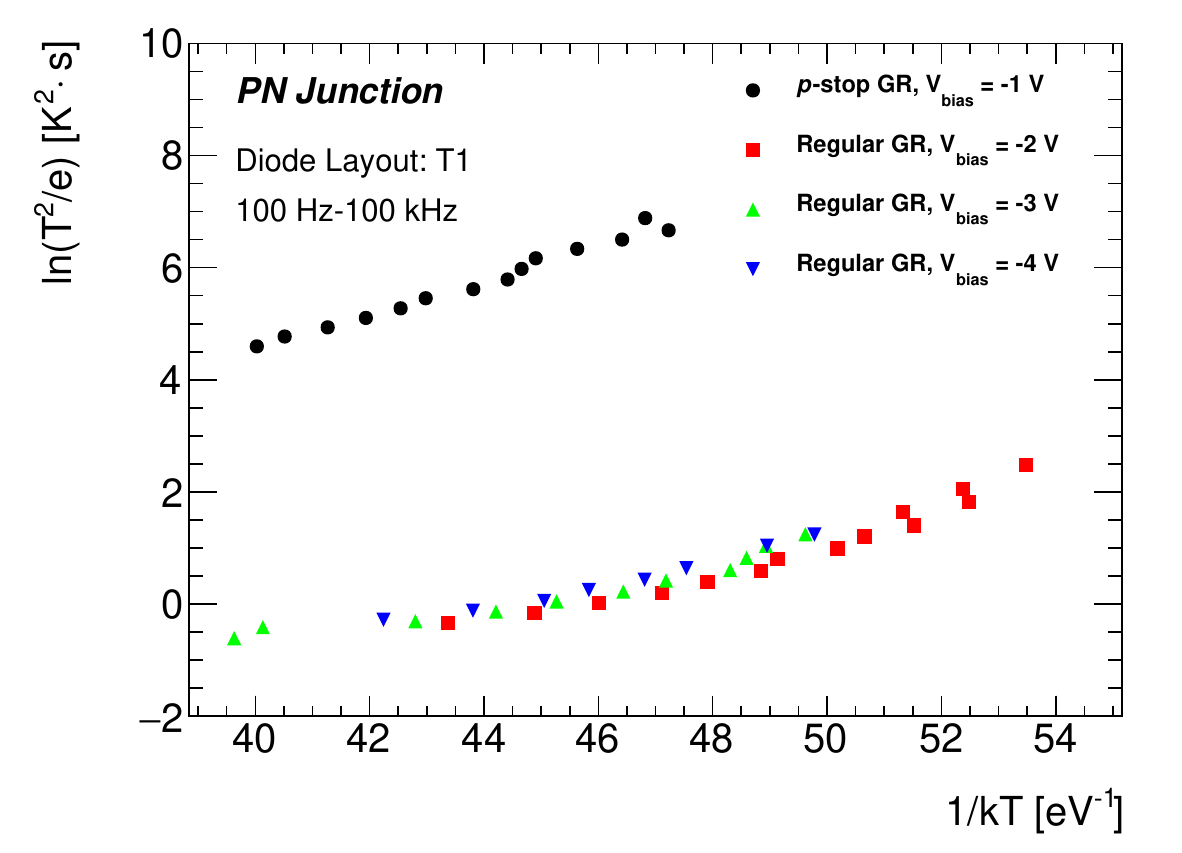}
\caption{Arrhenius plots obtained from TAS measurements of \textit{pn} junction diode samples.}
\label{fig:tas_pnjunction_arrhenius}
\end{figure}

In addition to standard DLTS measurements, TAS scans were performed on the samples.
This method not only complements the results obtained from DLTS, but also serves as a good method to test irradiated devices due to the limitations of DLTS for samples with high leakage current. 
The Arrhenius plots of \textit{pn} junctions with regular and \textit{p}-stop guard ring are shown in Figure \ref{fig:tas_pnjunction_arrhenius}, the obtained trap parameters are listed in Table \ref{table:tas_pnjunction}. 

\begin{table}[htbp]
\centering
\caption{Trap parameters obtained from TAS measurements for \textit{pn} junction diode samples.}
\label{table:tas_pnjunction}
\begin{tabular}{lccc}
\toprule
Sample & V\textsubscript{m} [V] & E\textsubscript{trap} [eV] & $\sigma$ [cm\textsuperscript{2}] \\
\midrule
\textit{p}-stop GR & $-$1 & 0.311$\pm$0.026 & 8.4$\times$10\textsuperscript{-19}$\pm$3.1$\times$10\textsuperscript{-19} \\ 
\midrule
\multirow{3}{*}{regular GR} & $-$2 & 0.277$\pm$0.043 & 1.0$\times$10\textsuperscript{-16}$\pm$8.3$\times$10\textsuperscript{-16} \\
 & $-$3 & 0.168$\pm$0.040 & 4.9$\times$10\textsuperscript{-19}$\pm$6.2$\times$10\textsuperscript{-19} \\
 & $-$4 & 0.209$\pm$0.037 & 3.3$\times$10\textsuperscript{-18}$\pm$5.5$\times$10\textsuperscript{-18} \\
\bottomrule
\end{tabular}
\end{table}

\subsubsection*{Schottky diode}
\label{subsubsection:schottky diode-dlts}

\begin{figure}[htbp]
\centering
\includegraphics[width=0.6\textwidth]{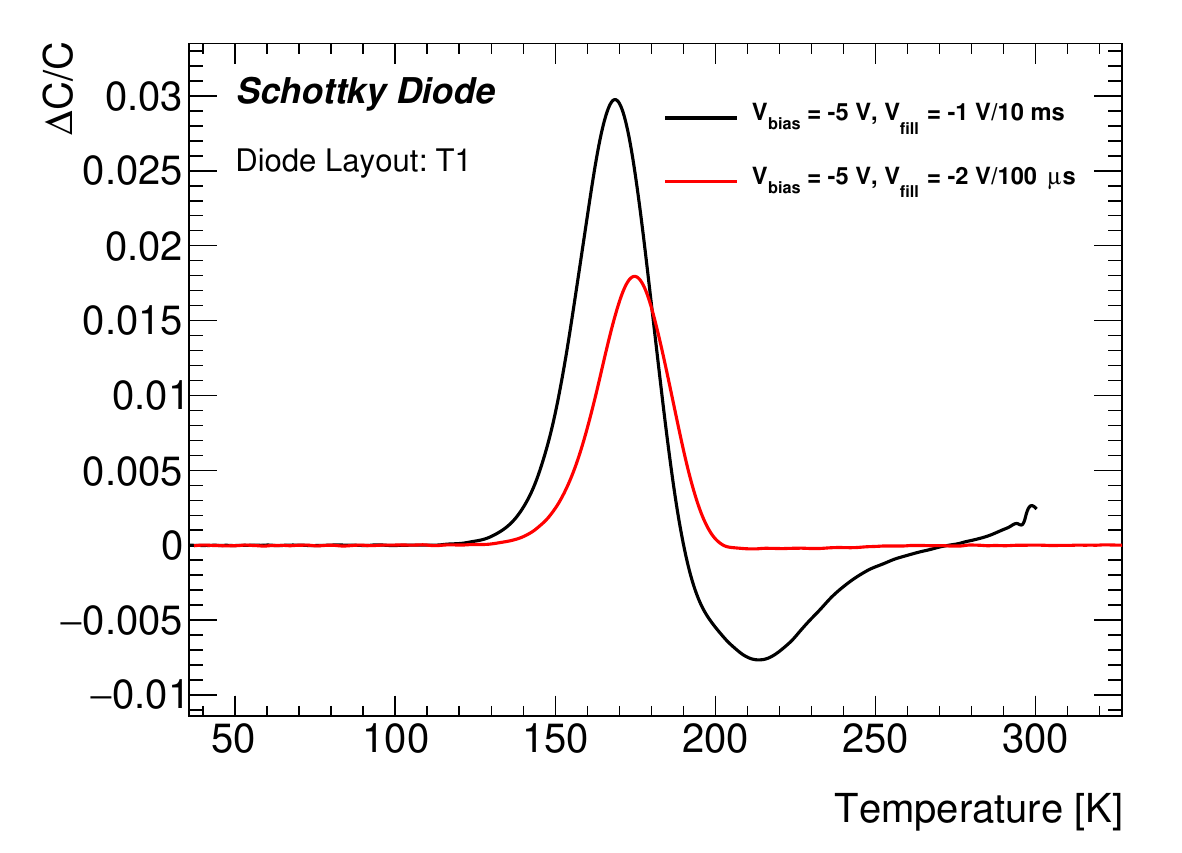}
\caption{DLTS spectra of a Schottky Diode for different measurement parameters.}
\label{fig:dlts_schottky}
\end{figure}

Similar to the \textit{pn} junction samples, the Schottky diodes have a common peak centered around 170 K.
However, as can be seen in Figure~\ref{fig:dlts_schottky}, unlike the \textit{pn} junction diodes no other defects were observed in DLTS measurements.
An additional minimum in the DLTS spectrum vanished when using shorter filling pulses, which indicates a small capture cross-section, and reduced voltages, thus avoiding surface and interface states from influencing the measurement.
The resulting trap parameters are summarized in Table~\ref{tab:dlts_schottky}.

\begin{table}[htbp]
\centering
\caption{Trap parameters obtained from DLTS measurements for Schottky diode samples.}
\label{tab:dlts_schottky}
\begin{tabular}{cccc}
\toprule
T\textsubscript{median} [K] & E\textsubscript{trap} [eV] & $\sigma$ [cm\textsuperscript{2}] & N\textsubscript{T}/N\textsubscript{s}\protect\footnotemark\\
\midrule
170 & 0.312 & 5.5$\times$10\textsuperscript{-15} & 7.8$\times$10\textsuperscript{-3} \\
180 & 0.294 & 3.3$\times$10\textsuperscript{-16} & 2.2$\times$10\textsuperscript{-2} \\
\bottomrule
\end{tabular}
\end{table}

\begin{figure}[htbp]
\centering
\includegraphics[width=0.6\textwidth]{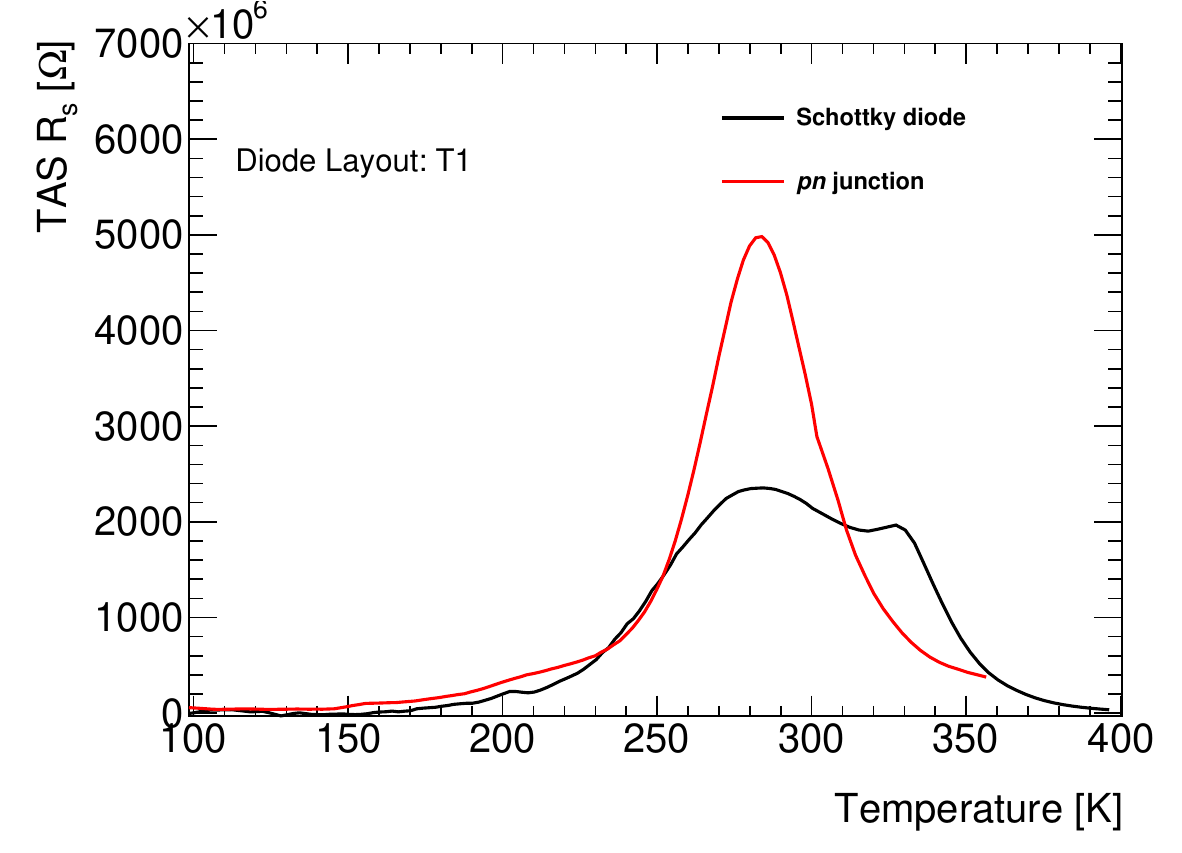}
\caption{TAS R\textsubscript{s} signal of a Schottky Diode (black) in comparison to a \textit{pn} junction diode (red) as a function of time for a fixed AC signal frequency.}
\label{fig:tas_schottky}
\end{figure}

TAS results for Schottky diodes show two trap states, while only one for \textit{pn} junction diodes, as seen in Figure~\ref{fig:tas_schottky}.
Arrhenius analyses of both (Table~\ref{tab:tas_schottky}) yield trap energy levels closer to the middle of the band gap than those obtained from standard DLTS.
In particular, the second trap near the midgap is expected to be a generation center and thus controlling the leakage current in the device.
\footnotetext{N\textsubscript{T}: the total density of deep level traps, N\textsubscript{s}: the total free charge carrier density.}

\begin{table}[htbp]
\centering
\caption{Trap parameters obtained from TAS measurements for Schottky diode samples.}
\label{tab:tas_schottky}
\begin{tabular}{cccc}
\toprule
Sample & V\textsubscript{m} [V] & E\textsubscript{trap} [eV] & $\sigma$ [cm\textsuperscript{2}] \\
\midrule
\multirow{4}{*}{Schottky diode} & \multirow{2}{*}{$-$1} & 0.498 & 1.6$\times$10\textsuperscript{-14} \\
& & 0.664 & 3.5$\times$10\textsuperscript{-13} \\ \cline{2-4}
& \multirow{2}{*}{2} & 0.467 & 3.0$\times$10\textsuperscript{-15} \\
& & 0.614 & 3.7$\times$10\textsuperscript{-14} \\
\bottomrule
\end{tabular}
\end{table}

\section{Conclusion and future plan}
\label{sectioin:conclusion and plan}
A research aimed at fabricating test structures, Schottky and \textit{pn} junctions on \textit{p}-type Silicon wafers, for the investigation of radiation bulk damage, has been described. The wafers used in this project are 6$''$ in size and feature a 50 $\mu$m thick epitaxial layer of five different levels of resistivity. 
The design consists of four different types of device layouts, differing in size to allow investigation of IV, CV and charge collection characteristics.
Currently, Schottky diodes on three wafers of the highest resistivities, corresponding to the doping concentration of $10^{13}$ cm\textsuperscript{-3}
have been successfully manufactured.
Another two wafers have been successfully used for the fabrication of \textit{pn} junctions, on wafers of doping concentration of $10^{13}$ cm\textsuperscript{-3}.
 Characterization of both Schottky and \textit{pn} junctions is underway. First test results of the performances of the non-irradiated devices have been shown, demonstrating good electrical characteristics of the fabricated devices and the reliability of the in-house fabrication process. 
Extrapolation of variables of interest, from reverse and forward mode of operation, from CV analysis and charge collection using a custom-made laser injection setup have been reported. Results of the defect investigations using the DLTS technique have also been shown.
The next step of this research is to complete the characterization by including results of neutron irradiated devices, from which we plan to extrapolate models to include in device simulators, and fabricate and test devices on wafers of medium (doping concentration: $10^{14}$ cm\textsuperscript{-3}) and low (doping concentration: $10^{15}$ cm\textsuperscript{-3}) resistivities.

\newpage
\appendix

\section{Device fabrication process}
\label{appendix:device fabrication process}

\subsection*{Schottky diode}
\label{subappendix:fabrication of the schottky diode devices}

\begin{figure}[htbp]
\centering
\includegraphics[width=0.8\textwidth]{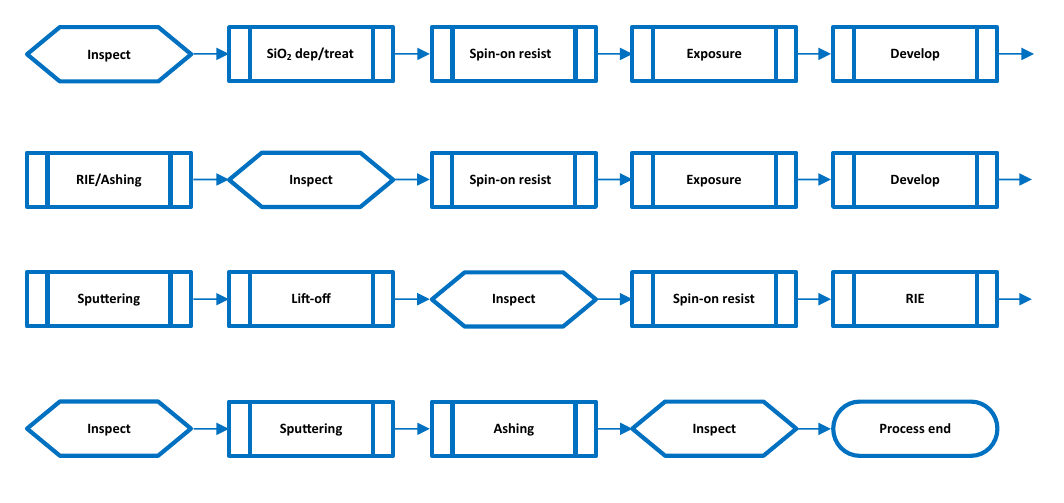}
\caption{Process flow of Schottky device fabrication.}
\label{fig:schottkyfabflow}
\end{figure}

The full synopsis of the manufacturing process is shown in Figure \ref{fig:schottkyfabflow}.
Firstly, a 500 nm oxide layer was deposited on the wafers via Low-Pressure Chemical Vapour Deposition (LPCVD). The oxide thickness was measured at several places across the wafer. Then a 1.5 $\mu$m thick layer of JSR-IX575 photoresist was spin-coated onto the wafer surface. The wafers were then given a soft bake and exposed using broadband UV light followed by a post exposure bake. Thereafter, the wafers were developed using the TMA238WA developer. After resist development the wafers were given a descum to remove any resist residues. Next, the oxide was etched in an Ar/CHF3 plasma using reactive ion etching. A white light reflectometer was used to check all the oxide had been etched. The resist was then stripped using acetone followed by an oxygen plasma in a barrel asher.
To define the metal contacts for the cathode and guard ring, a 3.3 $\mu$m thick layer of SPR220-7 photoresist was spun on the wafers followed by soft bake, a broadband UV exposure, subsequent hold time, post-exposure bake and development in MF26-A developer. This step was followed by a deposition of a 500 nm thick layer of Al using magnetron sputtering. The subsequent step of lift-off was performed in an ultrasonic bath of acetone. To aid resist removal the acetone was heated to 40 $^{\circ}$C via DI water and the wafers were subjected to regular ultrasonic bursts of 30 minutes. The LPCVD oxide was then etched from the backside of the wafer and a 1 $\mu$m thick layer of Al was deposited to create an Ohmic contact.
As the second process of photoresist deposition took place not immediately after the step of RIE oxide etching, with the wafers left exposed in free air inside the clean room, it is expected that a thin layer of native oxide grew over the top surface of the wafers. The presence of thin native oxide has been taken into account in the device analysis shown in Appendix \ref{appendix:schottky barrier in the presence of interface states}.

\subsection*{\textit{pn} junction}
\label{subappendix:fabrication of the pn junction diode devices}

\textit{pn} junction diode fabrication began with the growth of 150 nm of thermal oxide in a dry oxygen ambient at 1025 $^{\circ}$C. This oxide provides surface passivation and serves to mask the phosphorus diffusion used to form the junction. Oxidation was followed by an in-situ nitrogen ambient anneal for 20 minutes to reduce oxide fixed charge. Contact photolithography with wet etching was then used to open windows through the oxide for phosphorus diffusion to form large circular n\textsuperscript{+} cathodes. Oxide was left on the wafer backsides as a diffusion mask. Diffusion was carried out from a POCl3 vapour source diluted in nitrogen at 900 $^{\circ}$C for 5 minutes. Processing then split into three different paths for guard ring (GR) formation. For \textit{p}-stop GR samples photolithography with plasma etching was used to thin the oxide to 20 nm in annular rings surrounding the cathodes. 
$\prescript{11}{}{B}^{+}$ ions were implanted through the thinned oxide at $3\times 10^{13}$ cm\textsuperscript{2} dose and 20 keV energy to form a region of enhanced \textit{p}-type doping preventing surface inversion. Photoresist was left in place to mask the implant. The implant was annealed at 900 $^{\circ}$C for 10 minutes in nitrogen. For non-isolated GR samples oxide was completely removed in the annular ring, but these samples were not implanted. For isolated GR samples no additional processing was done at this stage. After etching 1\% hydrofluoric acid to hydrophobia to clear oxide from contact windows. A 750 nm aluminum layer was then deposited on the front of all samples by e-beam evaporation.  Photolithography with wet etching in hot phosphoric acid was used to pattern the aluminum. Swabbing with hydrofluoric acid was then used to remove oxide from the wafer backs, after which 500 nm of aluminum was deposited by e-beam to create a back contact. Contacts were sintered in pure hydrogen at 400 $^{\circ}$C for 10 minutes.

\section{Schottky barrier in the presence of interface states}
\label{appendix:schottky barrier in the presence of interface states}
Figure \ref{fig:psischottkybarrier} shows the band diagram of a Schottky junction on \textit{p}-type silicon with a thin insulating layer between the metal cathode and substrate:

\begin{figure}[htbp]
\centering
\includegraphics[width=0.6\textwidth]{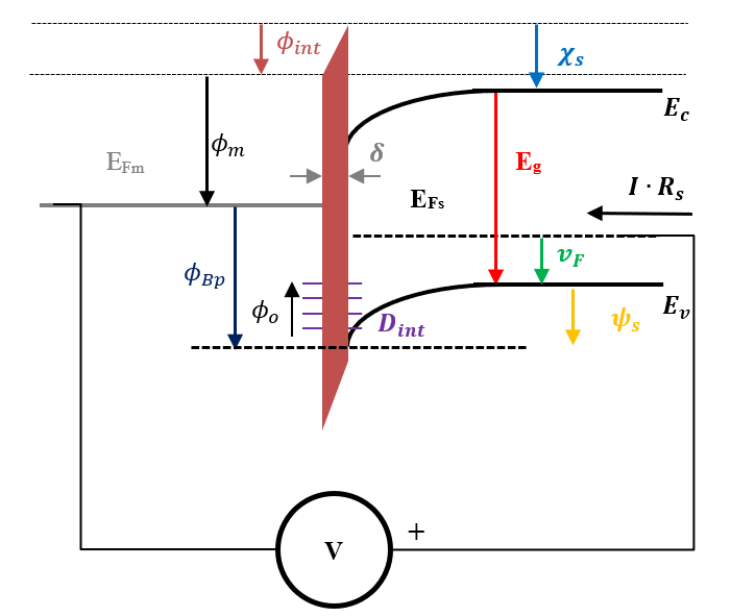}
\caption{Band diagram of Schottky junction on \textit{p}-type silicon, including a thin interface layer and interface states.
\label{fig:psischottkybarrier}}
\end{figure}

The voltage $\phi_{int}$ across the layer of thickness $\delta$ is given by:
\begin{equation}\label{SchBarrPhiIntI}
\phi_{int}=-V+IR_s-v_F-\psi_s+{\frac{1}{e}}(E_g+\chi_s-\phi_m) 	
\end{equation}
where $v_F$=$\frac{kT}{e}\ln{(\frac{N_v}{N_a})}$ [V] is the Fermi potential, $\psi_s$[V] is the surface potential, $E_g$ [eV] is the bandgap, $\chi_s$ [eV] the electron affinity of semiconductor, $\phi_m$[eV] the metal work function, $e$ [C] the elementary charge magnitude, $V$ the external voltage applied, $R_s$ the bulk resistance and $I$ the current flowing across the device.
From Gauss’s law:
\begin{equation}\label{GaussLaw}
\frac{\phi_{int}}{\delta}=\frac{Q_m}{\varepsilon_i} 
\end{equation}
where $Q_m$ is the charge accumulated on the metal side and $\varepsilon_i$ is the permittivity of the interface layer. The neutrality condition implies:
\begin{equation}\label{Qneutral}
Q_m =-Q_{sc}-Q_{ic} 
\end{equation}
where $Q_{sc}$ is the space charge in the semiconductor and $Q_{ic}$ is the interface charge. Under the full depletion approximation, furthermore:
\begin{equation}\label{Qsc}
Q_{sc}=-\sqrt{2e\epsilon_s N_a\psi_s}
\end{equation}
\begin{equation}\label{Qic}
Q_{ic}=-eD_{is}(\psi_s+v_F-\phi_o)							       \end{equation}
where $D_{is}$ [cm\textsuperscript{-2} eV\textsuperscript{-1}] is the density of interface states, $\phi_o$ [V] their neutral potential level from the top of valence band and $N_a$ [cm\textsuperscript{-3}] the dopant concentration.\\
From \ref{SchBarrPhiIntI}-\ref{Qic} one gets:
\begin{equation}\label{PsivsVimplicit}
-V+IR_s-v_F-\psi_s+{\frac{1}{e}(E}_g+\chi_s-\phi_m)=\frac{\delta}{\varepsilon_i}(\sqrt{2e\varepsilon_sN_a\psi_s}+\ eD_{is}\left(\psi_s+v_F-\phi_o\right)
\end{equation}

After collecting the terms, \ref{PsivsVimplicit} gives: 
\begin{equation}\label{PsivsVimplicitII}
\psi_s+\gamma\zeta{\sqrt\psi}_s=\gamma\frac{1}{e}\left(E_g+\chi_s-\phi_m\right)+\gamma\left(-V+IR_s\right)-v_F+(1-\gamma)\phi_o
\end{equation}

where $\gamma=\frac{\varepsilon_i}{\varepsilon_i+e\delta D_{is}}$ , $\zeta=\frac{\delta}{\varepsilon_i}(\sqrt{2e\varepsilon_sN_a})$.
\vspace{1cm}

Equation \ref{PsivsVimplicitII} expresses the non-linear relationship between $\psi_s$  and the applied voltage $V$. For no interface layer, i.e. $\gamma$=1 and $\zeta$=0, \ref{PsivsVimplicit} reduces to the Schottky-Mott limit~\cite{10.1063/1.321865,10.1007/BF01774216,mott_1938}, with the barrier height a function of the metal work function $\phi_m$ only. The presence of a thin layer reduces the Schottky-Mott and increases a Bardeen behavior of the junction~\cite{PhysRev.71.717}, introducing a non-linearity in the relationship through the second term in the left hand side of \ref{PsivsVimplicitII}. Furthermore, as the interface state density $D_{is}$ might depend on the energy and, thus, present different values depending on the voltage bias applied, the $\psi_s$ relationship with $V$ might take a rather complicated form. 
It is possible to simplify \ref{PsivsVimplicitII} in the case of an interface layer of small thickness, e.g. when only 1-2 nm of native oxide is present on the silicon surface~\cite{10.1063/1.347181,Bohling2016}. Indeed, the product $\gamma\zeta$, essentially a measure of the dominance of surface states charge over space charge, is usually < 1 up to relatively high values of doping $N_a$ and realistic values of $D_{is}$~\cite{Werner1994}, which allows neglecting the second term in left hand side of \ref{PsivsVimplicitII}, as shown in Figure \ref{fig:gammazetaprod}:

\begin{figure}[htbp]
\centering
\includegraphics[width=0.6\textwidth]{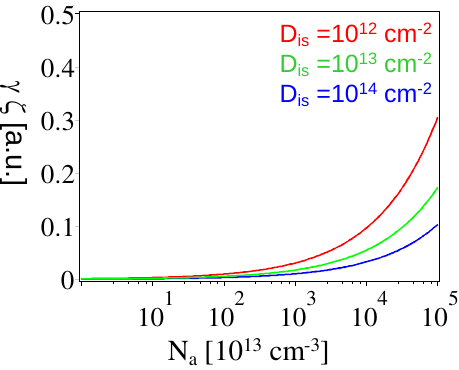}
\caption{The $\gamma\zeta$ product of equation \ref{PsivsVimplicitII} versus doping $N_a$ /10\textsuperscript{13} for different values of $D_{is}$ and $\delta$=2 nm.
\label{fig:gammazetaprod}}
\end{figure}

For doping $N_a$ up to $10^{17}$ cm\textsuperscript{-3}, \ref{PsivsVimplicit} can be approximated as:
\begin{equation}\label{PsivsV}
\psi_s=\gamma\frac{1}{e}\left(E_g+\chi_s-\phi_m\right)+\gamma\left(-V+IR_s\right)-v_F+(1-\gamma)\phi_o
\end{equation}

The expression \ref{PsivsV} for $V$ = 0 gives the barrier height at zero bias $\phi_{Bp0}$:

\begin{equation}\label{ZeroBarHeight}
\psi_{s0}+v_F=\phi_{Bp0}=\gamma\frac{1}{e}\left(E_g+\chi_s-\phi_m\right)+(1-\gamma)\phi_o
\end{equation} 			        

\section{Schottky barrier IV characteristics in the presence of interface states}
\label{appendix:schottky barrier iv characteristics in the presence of interface states}
To have good rectifying properties, a Schottky diode should be fabricated using metals of suitable work function, to guarantee high barrier height to the semiconductor substrate.
According to the Schottky-Mott model the barrier height $\phi_{bp}$ of an ideal contact between a metal and a \textit{p}-type semiconductor in the absence of surface states depends on the difference between the metal work function $\phi_{m}$ and the electron affinity $\chi_{s}$ of the semiconductor:

\begin{equation}\label{SchottkyBarrierGenForm}
\phi_{bp}=\frac{1}{e}(E_g+\chi_s-\phi_m) 	
\end{equation} 			        
where e[C] is the elementary electron charge and $E_g$[eV] is the band gap energy. Choosing the correct metal is therefore essential to obtain a high barrier height~\cite{Li2006}. As aluminium has a relatively low work function~\cite{SMITH197171}, it is a suitable candidate for fabricating Schottky devices on \textit{p}-type silicon.
However, various reports~\cite{CankayaUcar+2004+795+798,SCHLESINGER19951,HARA1997394} show that the barrier height $\phi_{bp}$ indeed changes with $\phi_m$ but does not scale linearly as predicted by the Schottky-Mott model \ref{SchottkyBarrierGenForm}. The model proposed by Bardeen includes the presence of interface states, due to silicon dangling bonds at the surface. These states might cause a pinning of the Fermi level which influences the barrier height. The resulting relationship between metal work function and barrier height in the presence of a thin insulating layer and related interface states is given by \ref{PsivsVimplicitII}.
With the doping concentrations satisfied in this project, the model of Cowley and Sze~\cite{10.1063/1.1702952,Mander1982PhysicsOS}, which essentially connects the Schottky-Mott and Bardeen models, is however valid and is expressed by \ref{PsivsV}.

The parameters of the interface can be determined from the forward biased diode characteristics~\cite{10.1063/1.97359}, which, under the assumption of thermionic emission and neglecting image force lowering~\cite{VanderielAldert,Li2006} express the current as:

\begin{equation}\label{FWD_Sch_Current_GenForm}
  I=S A^{*} T^2 e^{(-\frac{e \phi_{Bp0} }{kT})} e^{(\frac{eV}{nkT })}  =I_0 e^ {(\frac{e}{nkT}(V-RI))}  
\end{equation} 			        
where $S$ [cm\textsuperscript{2}] is area of the device, $A^*$ [A$\cdot$cm\textsuperscript{-2}$\cdot$K\textsuperscript{-2}] the Richardson’s constant, around 32  for \textit{p}-type silicon, $T$ [K] the temperature, $n$ the ideality factor, $\phi_{Bp_0}$ [eV] the barrier height between semiconductor and metal at zero bias and $R$ is the resistivity of the bulk, which causes an ohmic drop in the applied voltage. From \ref{FWD_Sch_Current_GenForm}, by extrapolation at $V=0$, one obtains:
\begin{equation}\label{PhiBp0fromIV}
 \phi_{Bp0}=\frac{kT}{e}ln(\frac{SA^{*}T^2}{I_{V=0}})  
\end{equation} 	
and, after some algebraic manipulation:
\begin{equation}\label{FWD_dlnI_dV}
  \frac{dV}{dln(I)}=\frac{nKT}{e}+RI  
\end{equation} 			        
which, plotted as function of $I$, provides R as the slope in the linear region of the plot and values of $n(I)$, from which the ideality factor $n(V)$ can be obtained. 
Assuming the interface states equilibrate with the semiconductor only~\cite{HCCard_1971}, the ideality factor $n$ can be expressed as:
\begin{equation}\label{n_idel_vs_Dis}
 n(V)=1+\frac{\delta}{\epsilon_i}(\frac{\epsilon_{si}}{W}+eD_{is})
\end{equation} 			        
where $\epsilon_i$ is the permittivity of the insulating layer between silicon and the metal, $\epsilon_{si}$ the permittivity of silicon, $W$ is the width of the space charge region and $D_{is}$ is the density of interface states. From \ref{n_idel_vs_Dis} one gets:
\begin{equation}\label{Dis_vs_n}
 D_{is}(V)=\frac{1}{e}(\frac{\epsilon_i}{\delta}(n(V)-1)-\frac{\epsilon_{si}}{W})
\end{equation}

In the case of a \textit{p}-type silicon, the interface state energy $E_s$  measured from the top of the valence band $E_v$  is given by:
\begin{equation}\label{Es_vs_V}
 E_s-E_v=e(\phi_{Bp0}-\frac{V}{n(V)})
\end{equation} 			        

Using \ref{n_idel_vs_Dis}, \ref{Dis_vs_n}, the plot of interface states $D_{is}$ versus energy $E_s-E_v$ can be obtained. From \ref{ZeroBarHeight} and the value of $D_s$ at $V$=0 the value of neutrality level $\phi_0$ can also be obtained.

\section{Schottky capacitance in the presence of interface states}
\label{appendix:schottky capacitance in presence of interface states}

Using \ref{Qsc} and \ref{Qic}, the capacitance per unit area is given by:
\begin{equation}\label{CapGenForm}
C=\frac{d}{dV}(Q_{sc}+Q_{ic})=\frac{d}{d\psi_s}(Q_{sc}+Q_{ic})\frac{d\psi_s}{dV} 	
\end{equation}

The full expression of \ref{CapGenForm} would require using \ref{PsivsVimplicitII} and current expression \ref{FWD_Sch_Current_GenForm} to calculate the derivative of $\psi_s$ versus $V$.
However, using the approximate expression \ref{PsivsV}, one gets\footnote{The derivative $\frac{d\gamma}{dV}$ is assumed to give negligible contribution.}:
\begin{equation}\label{CapFormV}
C\left(V\right)=\left(\sqrt{\frac{e\varepsilon_sN_a}{2\psi_s}}+eD_{is}\right)\frac{\gamma}{1+IR_s\frac{e}{n\ kT}}\cong\left(\sqrt{\frac{e\varepsilon_sN_a}{2\psi_s}}+eD_{is}\right)\frac{\gamma}{1+\gamma I R_s\frac{e}{\ kT}} 
\end{equation} 		

The approximation of $\frac{1}{\ n}= \gamma$ in the right hand side of \ref{CapFormV} stems from \ref{n_idel_vs_Dis}, assuming negligible\footnote{The extension of depletion width $W(V)$ renders the second term in right hand side of \ref{n_idel_vs_Dis} negligible w.r.t. the first. This is certainly verified for low doping conditions and up to moderate forward biasing.} the contribution of $\frac{\varepsilon_{Si}}{W\left(V\right)}$ to $D_{is}$. Expression \ref{CapFormV} shows that the capacitance depends non-trivially on the interface states. However, in the reverse biased region and for small leakage current $I$, the contribution in the denominator of second term in right hand side of \ref{CapFormV} is negligible\footnote{In case of irradiated devices the condition of low leakage might not be fulfilled, a case not treated here. This would require a different approach to the calculation of capacitance, due to the introduced defects in the bandgap.}, from which:
\begin{equation}\label{CapFormVapprox}
C\left(V\right)\cong\left(\sqrt{\frac{e\varepsilon_sN_a}{2\psi_s}}+eD_s\right)\ \gamma 
\end{equation} 

The effect of interface states on capacitance is expected to decrease as the frequency of the AC signal applied is increased, owing to the inability of the states to follow rapidly changing signals. Thus, at high frequency, and neglecting the second term in parenthesis in the right hand side of \ref{CapFormVapprox}, one gets:
\begin{equation}\label{Cap2FormV}
C^{-2}\left(V\right)\cong\frac{2\psi_s}{e\varepsilon_sN_a}\gamma^{-2}
\end{equation} 

Substituting in \ref{Cap2FormV} expression \ref{PsivsV} for $\psi_s$ and equating to 0 gives for the diffusion potential $\psi_{s0}$:
\begin{equation}\label{DiffPotForm}
V_0 - I_0R_s=\frac{\frac{\gamma}{e}\left(E_g+\chi_s-\phi_m\right)-v_F+(1-\gamma)\phi_o}{\gamma}=\frac{\psi_{s0}}{\gamma}
\end{equation} 
i.e. the point of intercept of $C^{-2}$ with the $V$ axis provides the $\gamma$-scaled diffusion potential $\psi_{s0}$.

From \ref{Cap2FormV}, the slope of $C^{-2}\left(V\right)$ with respect to $V$ is given by:
\begin{equation}\label{DCap2FormV}
\frac{dC^{-2}\left(V\right)}{dV}\cong\frac{2}{e\varepsilon_sN_a}\gamma^{-1}
\end{equation} 
i.e. the slope depends not only on the actual doping $N_a$ but also on the density of interface states via $\gamma^{-1}$.

\acknowledgments
This work is supported by the European Organization for Nuclear Research (CERN) under grant RD50-2019-03, the Science and Technology Facilities Council (STFC) of United Kingdom under grants ST/S000747/1 and ST/W000547/1, National Science and Engineering Research Council (NSERC) of Canada under grants SAPPJ-2022-00020 and RTI-2020-00105.
We are also grateful for the support with equipments and technical personnel by the Rutherford Appleton Laboratory, the Department of Physics and Astronomy of University of Sheffield, the Department of Electronics of Carleton University, the Institute of High Energy Physics of Chinese Academy of Sciences, the University of Chinese Academy of Sciences and the School of Physics of Zhejiang University. The authors would like to thank Dr. Dan Johnstone of Semetrol LLC for many useful insights and initial sample cross-checking for the DLTS measurements.
The authors would also like to thank Dr. Ioana Pintilie of the National Institute of Material Physics, Bucharest-Magurele, Romania, for her DLTS analysis of samples.

\clearpage
\bibliographystyle{JHEP}
\bibliography{biblio.bib}

\end{document}